%% file: main.tex
\documentclass[fleqn,10pt]{wlscirep}
\usepackage[english]{babel}
\usepackage[utf8]{inputenc}
\usepackage[T1]{fontenc}
\usepackage{bm}
\usepackage{tikz}
\usepackage{todonotes}
\usepackage{xcolor}
\usepackage{pgffor}
\usepackage{ifthen,xstring}
\usepackage{bbold}

\usepackage[normalem]{ulem}

\usetikzlibrary{calc}
\usetikzlibrary{external}

\input{tikzstuff} 

\title{Tensor Networks for Quantum Machine Learning}
\author[1,*]{Hans-Martin Rieser}
\author[1]{Frank Köster}
\author[1,+]{Arne Peter Raulf}
\affil[1]{Deutsches Zentrum für Luft- und Raumfahrt, Institute for AI safety and security, Ulm / St. Augustin, Germany}

\affil[*]{e-mail: hans-martin.rieser@dlr.de, https://orcid.org/0000-0002-1921-1436}
\affil[+]{https://orcid.org/0009-0003-8672-3014}

\begin{abstract}
Once developed for quantum theory, tensor networks have been established as a successful machine learning paradigm. Now, they have been ported back to the quantum realm in the emerging field of quantum machine learning to assess problems that classical computers are unable to solve efficiently. Their nature at the interface between physics and machine learning makes tensor networks easily deployable on quantum computers. In this review article, we shed light on one of the major architectures considered to be predestined for variational quantum machine learning. In particular, we discuss how layouts like MPS, PEPS, TTNs and MERA can be mapped to a quantum computer, how they can be used for machine learning and data encoding and which implementation techniques improve their performance.

\end{abstract}
\begin{document}

\flushbottom
\maketitle

\thispagestyle{empty}

\section{Introduction}

Quantum computation is widely believed to set a new paradigm in computation. Utilizing quantum phenomena allows to solve certain problems
\cite{Shor.1994} far more efficient than classical binary algorithms. This raises hope that quantum implementations of other tasks also may provide quantum advantages.

One of the applications that could benefit from the access to the high dimensional Hilbert spaces of quantum computers is machine learning (ML). ML is a data driven approach for solving complex problems. An ML algorithm generates a model from training data that can be used to make predictions against previously unseen data. Quantum machine learning (QML) could advance learning by improved generalization to unknown data~\cite{Caro.2022}, higher noise robustness and the need for less training data~\cite{Abbas.2021}, and provide a more natural approach to quantum data analysis circumventing intermediate measurements~\cite{Perrier.15.08.2021} or generally a better computational complexity scaling~\cite{Boixo.2018}. 

Promising candidates for QML architectures are tensor networks (TN). They provide a structured approach for handling large objects with tensor structure which carry high amounts of correlated information like quantum states. Initially developed to store and process physical states of many-body quantum systems in numerical simulations\cite{White.1992,StellanOstlund.1995}, TNs also turned out to be useful for ML applications.
Their approach to realize learning architectures is complementary to neural networks. 
As the TN description uses a (quantum) state and operator formulation, the transfer to a quantum computer can be done naturally. 

In this review, we focus on the application of TNs for QML. We will begin with a short introduction to the classical TN theory including optimization and ML approaches in Section~\ref{sec:class}. Then, we will discuss how to apply these concepts to a quantum computer in Section~\ref{sec:qtn} and the encoding of data to quantum states for ML applications in Section~\ref{sec:Encoding}.

We will not cover many aspects of classical TNs in detail. For a deeper technical dive into TNs, the reader may refer to a general introduction~\cite{Bridgeman.2017} and the reviews on specific layouts~\cite{Cirac.2021,Evenbly.2009} or decomposition and optimization techniques~\cite{Schollwock.2011,Cichocki.2016,Vanderstraeten.2019}. Applications are many-body quantum systems~\cite{Orus.2019}, nonlinear system identification~\cite{Batselier.2022} and classical ML~\cite{Cichocki.2016b,Levine.2023}.

The field of TN-QML is just developing, and notations and terminology vary throughout the community. Due to their origin in quantum theory, some authors call even ML with classical TNs "quantum machine learning"\cite{Liu.15.05.2020}. In our opinion, a more suitable term would be \emph{quantum-inspired} here. Furthermore, one can argue that variational quantum circuits (VQC)~\cite{Schuld.2021} require classical optimization and therefore are hybrid. In this article however, we will use the following convention: Methods fully evaluated without a quantum computer will be called \emph{classical}. Methods developed for quantum computers that only require classical optimization of weights will be called \emph{quantum} TNs (QTN), as full quantum computation is still far out of reach. The term \emph{hybrid} will be used for methods that combine QML with a classical data processing structure, e.g. pre-training or data pre- and post-processing.

\section{Classical Tensor Networks}\label{sec:class}
    
\subsection{Introduction on Tensors and Tensor Networks}\label{ssec:tensor}

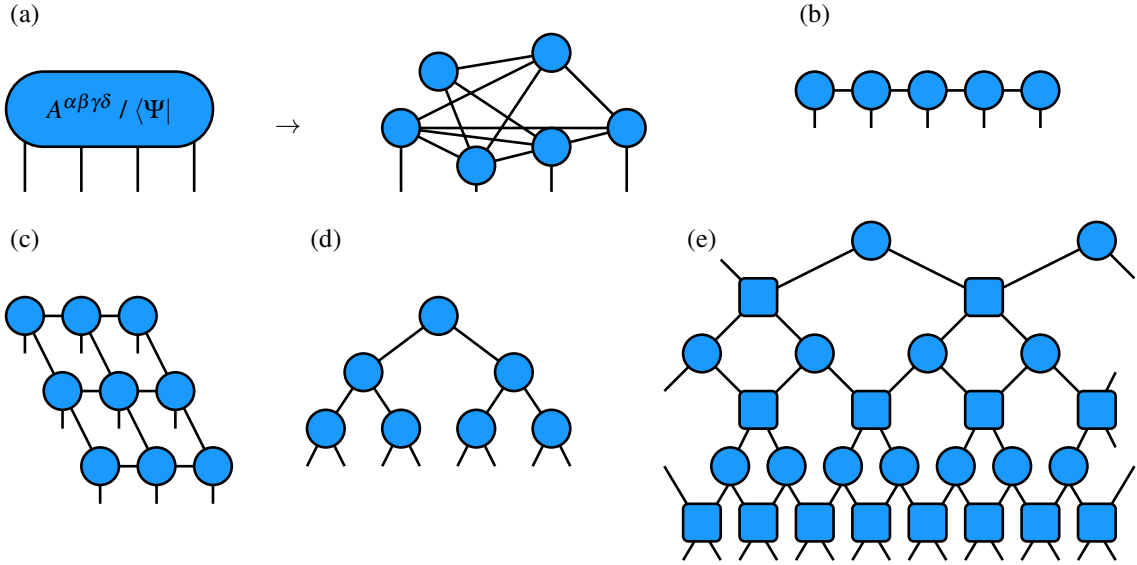
\begin{figure}
	    \centering
	    \input{structure}
	
	\caption[Examples for common tensor network layouts]{Examples for common tensor network layouts. (a) a general irregular tensor network. One may use any tensor network structure to express the large tensor $A^{abcd}$ or the wave function $\langle\Psi|$. However, regular tensor networks provide benefits in terms of interpretability and universality. Both (b) matrix product states (MPS) and (c) projected entangled pair states (PEPS) share the same grid structure with different dimensionality. (d) tree tensor networks (TTN) and (e) multiscale entanglement renormalization ansatz (MERA) have a hierarchical structure, where MERA entangle between individual branches in contrast to TNN. 
    }
	    \label{fig:structure}
	\end{figure}

TNs are a decomposition of large tensorial structures into several connected low rank tensors (see Figure \ref{fig:structure}~(a)). Tensors are multidimensional arrays and therefore generalizations of vectors and matrices. While a matrix has two indices a tensor may have an arbitrary amount of indices. Technically, tensors describe objects from and maps between tangent and cotangent spaces. A tensor may have regular (lower) and dual (upper) indices depending on whether this index refers to objects from a tangent space or from a cotangent space. Each of these spaces may have different dimensions. A single tensor may have both types of indices and therefore connect to both tangent and cotangent spaces. The rank of a tensor corresponds to its number of free indices.

The electromagnetic field tensor $F_{ab}$ from relativistic physics for example is a rank two tensor with four dimensional space time indices each and the Riemann curvature tensor from general relativity ${R^a}_{bcd}$ is a rank four object, with one dual ($a$) and three regular ($b$, $c$, $d$) indices. Free regular indices can be contracted with free dual indices by summing over all dimensions of this index. Einstein sum convention is a convenient form to express this contraction: having the same index twice automatically implies a summation

    \begin{equation}
        -\frac{1}{4\mu_0}F_{ab}F^{ab}=-\frac{1}{4\mu_0}\sum_{a,b=0}^{3}F_{ab}F^{ab}.
    \end{equation}

As writing these tensors with indices can be very complex for larger problems, graphical notations like the Penrose diagrams have been developed to simplify the handling of tensor equations\cite{Schollwock.2011}. The tensors from before correspond to the diagrams 

    
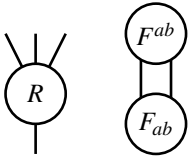
\begin{tikzpicture}[scale=.8]
        \draw[line width = 1pt] (0,-1)--++(0,2);
        \draw[line width = 1pt] (-.5,1)--(0,0)--(.5,1);
        \draw[line width = 1pt, fill=white] (0,0)circle(.5);
        
        \draw[line width = 1pt] (1.75,-.5)--++(0,1.5);
        \draw[line width = 1pt] (2.25,-.5)--++(0,1.5);
        \draw[line width = 1pt, fill=white] (2,1)circle(.5);
        \draw[line width = 1pt, fill=white] (2,-.5)circle(.5);
        \node at (0,0){$R$};
        \node at (2,1){$F^{ab}$};
        \node at (2,-.5){$F_{ab}$};
    \end{tikzpicture}

The graphical notation actually is one strength of the TN paradigm, as it provides accessibility to high dimensional states: Each symbol is a tensor, its rank is given by the number of legs it has and the type of index determines the direction of the associated leg. 

Figure \ref{fig:structure}~(a) illustrates the idea behind the TN approach: A large tensor $A^{abcd}$ which may represent some quantum state $\langle\Psi|$ usually is hard to handle computationally. It requires large storage space and the manipulation of a large number of entries for each operation. Breaking down $A$ into a network of smaller connected tensors improves computability when the internal structure of $A$ matches the TN's layout. This requires a third kind of tensor index called \emph{internal} or virtual index that connects the constituents of the TN. We will denote this kind of index in greek letters. The dimension of internal indices is called bond dimension $\chi$. It determines how strongly the constituent tensors are coupled and how much information is shared between them.

TNs allow to apply local operations individually on each tensor node instead of having to evaluate the whole tensor at once. Tensors can be joined by contracting over connected indices or decomposed into several connected tensors. The most common technique for decompositions along a single direction is singular value decomposition (SVD), a generalization of diagonalization for arbitrary shaped tensors. Polar decomposition is faster than SVD, but does not allow for reducing bond dimensions easily. Tucker decomposition can be used for decomposing nodes within several directions at once\cite{TamaraG.KoldaandBrettW.BaderSandiaNationalLaboratories.2018}. 

The general idea behind tensor decomposition methods is to represent an arbitrary tensor with a specific set of constituent tensors. In SVD for instance, a tensor $A^\alpha_{a\delta}$ is decomposed into a unitary matrix $U^\alpha_\beta$, a diagonal singular value matrix $\Sigma^\beta_\gamma$ and an isometric matrix $V^\gamma_{a\delta}$

\begin{equation}
    \begin{tikzpicture}[baseline=(current  bounding  box.center)]
        \node[left] at (-.6,0){$A^\alpha_{a\delta} =$};
        \draw[line width=1pt](-.5,0)--++(1,0);
        \draw[line width=1pt](0,0)--++(0,.45);
        \TNode{}{0,0}{white}{};
        \node at (.75,0) {$=$};
        \node at (.75,.25) {\tiny SVD};
        \draw[line width =1pt](1,0)--(3.35,0);
        \draw[line width =1pt](3,0)--++(0,.45);
        \TNode{C}{1.5,0}{white}{$U$};
        \TNode{}{2.25,0}{white}{$\Sigma$};
        \TNode{2}{3,0}{white}{$V$};
        \node[right]at(3.5,0){$=U^\alpha_\beta\Sigma^\beta_\gamma V^\gamma_{a\delta}$};
    \end{tikzpicture}
\end{equation}

where isometric tensors with known direction are given by triangles, unitaries and isometries with unknown orientation by squares and any other kind of tensor by a circle. Having access to the singular values in the diagonal matrix $\Sigma$ allows for reducing bond dimensions by removing zero singular values. This also can be used for approximation removing the lowest singular values having the least contribution to the bond.

Since tensor decompositions can be done in any direction on each bond and contracted to each side at any time, TNs are not unique but contain a gauge degree of freedom. One can make use of this property to bring the TN to a canonical form where the bonds form orthonormal Hilbert spaces\cite{Orus.2019} and the tensors are isometric or even unitary\cite{Bridgeman.2017}. In many cases, it makes sense to bring the TN to such a canonical form where all tensor nodes are isometric. This has several advantages. First of all, isometric tensors automatically fulfil a normalization condition $A^{a\mu} A^\dag_{a\Tilde{\mu}} = \delta^\mu_{\Tilde{\mu}}$ which enables the application of optimization schemes (see Section \ref{ssec:opt}). Second, it is mandatory for techniques that require directionality \cite{Zaletel.2020} or make use of the properties of isometries \cite{Geng.2022}. In particular, mapping a TN to a quantum circuit requires the tensors to be at least isometric (see Section \ref{sec:qtn}). 


\paragraph{Applications in Quantum Computing} are based on the original appliction of TNs: reducing the computational cost of storing and evaluating lowly entangled multi-particle quantum states. This comes in handy for quantum computer simulations both for execution\cite{Zhou.2020,Nguyen.21.04.2021} and validation\cite{McCaskey.2018} of circuits as well as the estimation of errors\cite{Guo.2020}. Especially for short NISQ era algorithms, entanglement between many qubits usually is not too high and therefore circuit sizes well beyond the power of other simulation methods can be evaluated using TNs\cite{Pednault.16.10.2017}. 

Additionally, TNs have been proposed to parallelize quantum simulations by cutting the system into several weakly entangled pieces and approximating the state of all but one piece by TNs\cite{Barratt.2021}. Simulating a quantum computer may indeed be more resource efficient than using quantum hardware itself for a lot of low-entanglement applications\cite{Jaschke.24.05.2022}. This idea has been used already to develop quantum inspired algorithms executed on classical hardware, e.g. for optimizing stock market portfolios\cite{Alcazar.15.01.2021, Mugel.2022} or radiotherapy plans\cite{Cavinato.2021} with quantum algorithms compressed to a classical TN approximation.

\subsection{Tensor Network Layouts}\label{ssec:layout}

Technically, the TN may have any shape but using regular TNs provides many benefits like simpler optimization, simpler control and transferability to problems with different structure. Such TNs are also more interpretable than arbitrary networks. The most common layouts either are grid (Fig. \ref{fig:structure} b and c) or hierarchical (Fig. \ref{fig:structure} d and e) states. Promoting state layouts to operators is either done by allowing every individual grid tensor node to have regular and dual indices or by connecting a complete hierarchical network with its dual on their topmost layers. 

\paragraph{Grid Layouts} are the most natural TN description of physical lattices as the layout has a similar structure to the system. These layouts can be seen as derivatives of Projected Entangled Pair States (PEPS)\cite{Sierra.1998}. In quantum applications, PEPS nodes are constructed as composite objects consisting of coupled internal spins. Each spin connects to a neighboring site via an edge and at each node the constituent spins are entangled and truncated, thus the name PEPS. The number of spin tuples depends on the the dimensionality of the network\cite{Cirac.2021}, typically a hypercube or hexagonal.

The constituent spin construction is very useful when employing PEPS for the description of quantum systems as this allows for spin constraints on the bonds. For ML applications however, ansaetze for the nodes reflect computational approximations or inductive biases. 

Although PEPS are defined for arbitrary dimensions, usually low dimensional layouts are used. One dimensional PEPS are called Matrix Product States (MPS) or tensor trains. 
These are the simplest and most studied TN layouts\cite{Cirac.2021}. In index notation, the MPS from Fig.~\ref{fig:structure}~(b) will look like

    \begin{equation}
    	A^{abcde} = \tilde{A}^{a(1)}_{\alpha}\;\tilde{A}^{\alpha b(2)}_\beta\;\tilde{A}^{\beta c(3)}_\gamma\;\tilde{A}^{\gamma d(4)}_\delta\;\tilde{A}^{\delta e(5)}
    \end{equation}
with constituent tensors $A^{(k)}$. Common gauges for MPS are called left, right and site canonical forms depending on the orientation of the isometric tensor nodes\cite{Cirac.2021}.

Brickwall or checkerboard TNs used in some quantum computing applications\cite{Uvarov.2020, Lubasch.2020, Lazzarin.2022} are another variety of two dimensional grid layouts equivalent to a hexagonal PEPS. The brickwall layout is a superposition of MPS up to a certain bond dimension\cite{Lubasch.2020} as it allows for the realization of MPS of different gauges overlapping at the same time.

\paragraph{Hierarchical Layouts} have input or output tensor nodes that are not coupled directly but are pooled on several internal layers. The simplest hierarchical structure is a tree tensor network (TTN) where two or more child nodes are connected to a parent node in the next layer until only a single node is left on the top. This layout is also called hierarchical Tucker decomposition. TTNs are able to catch both local entanglement and long range entanglement between groups of nodes, but not long range entanglement between individual tensor nodes. 
A TTN may have variable depth on different branches when the considered system is not homogeneous\cite{Murg.2015}.

The Multi Scale Entanglement Renormalization Ansatz (MERA) is an isometric TTN derivative with better entropy scaling \cite{Vidal.}. The main idea is to enhance the hierarchy with layers of unitary nodes connecting neighboring branches. These so called \emph{disentanglers} reduce entanglement passed on to the next level (see figure \ref{fig:structure} e).
MERA has a higher computational cost than other layouts due to the loops, but it can capture symmetry and far higher entanglement\cite{Cirac.2021,Araz.21.02.2022} while still being efficiently storable\cite{Bridgeman.2017}. Varieties of MERA 
offer even better entropy scaling \cite{Orus.2019}. Both TTN and MERA can be generalized to higher dimensions by considering unit cells of the respective dimension at each node\cite{Tagliacozzo.2009,Cincio.2008}. 

\vspace{1em}
    
The layout of a TN determines the maximal entanglement or internal correlation it can  support. This gives a bound on the system type the TN can approximate without having a bond dimension scaling exponentially with the system size. For MPS and PEPS entanglement fulfils an area law which means, that the amount of entanglement between a sub-network and its surroundings scales with
its boundary\cite{Wolf.2008}. This means, the entanglement for an 1-D MPS is constant \cite{Cirac.2021}, for a 2-D PEPS it scales linearily. For MERA based layouts, the entaglement scales up to a volume law, where a sub-networks entanglement with the surroundings depends on the number of nodes within the sub-network.

In practice, the choice for a specific layout usually is a trade-off between the possible entanglement and the computational cost: MPS and TTN can be contracted efficiently, MERA and PEPS usually are costly.

Further refinements can be made by applying symmetries to the TN\cite{Cirac.2021}. Relevant symmetric systems are homogeneous or periodic grids or layers in hierarchical networks\cite{Bridgeman.2017}. For ML, this reduces the complexity of the TN and makes it easier to train. 

\subsection{Optimization Methods}\label{ssec:opt}

The term 'optimizing TNs' can refer to two things. The size of a TN representation can be reduced by iterative executions of tensor decompositions along the internal bonds. This allows for the local adaption of bond dimensions to relevant degrees of freedom, e.g. by defining a threshold for relevant singular values. 

More often however, one seeks to optimize the value of some function of the TN. In quantum physics for example, this means to maximize the overlap between some given state and a TN approximation or minimizing the energy expectation value with respect to some Hamiltonian to find its TN ground state. This corresponds to minimizing a loss function of a TN based ML approach. The optimization can be achieved via several well established methods. In particular, general global gradient methods are available as well as TN specific techniques which make use of the network's locality and the tensorial nature of the nodes.

\paragraph{Renormalization methods} make use of the gauge ambiguity in TNs. They exploit the locality of operators to optimize the TN site by site. Density matrix renormalization group (DMRG), the first method of this kind, was developed to optimize spin chain Hamiltonians efficiently \cite{White.1992}. Soon, it was understood that restricting the maximum entanglement at each site reduces computational resources while describing lowly-entangled chains very well\cite{Vidal.2004} and further renormalization techniques were developed\cite{Daley.2004}. These provide powerful tools for optimizing MPS. 
Renormalization methods for TNs have been reviewed extensively before\cite{Schollwock.2011,Bridgeman.2017,Orus.2019}. Hence, we will only sketch the basic idea of DMRG for a finite MPS here.

DMRG can be applied to Hamiltonians $H$ that consist of independent blocks connecting neighboring MPS nodes. First, initialize a state randomly and consider the expectation value $\langle\Psi_0 | H |\Psi_0\rangle$. Start with a block at one end of the chain and contract all other nodes to an environment tensor generating an effective Hamiltonian for the first site. Diagonalize the effective Hamiltonian and truncate its Hilbert space to the lowest (effective) eigenvalues. Subsequently iterating this procedure at each site, will deterministically evolve the MPS towards the Hamiltonian's groundstate.

Renormalization methods provide a local and fast way of optimization adapted to the structure of TNs but also have some disadvantages. First, DMRG is hard to implement in standard ML frameworks, especially when combining TNs and neuronal layers\cite{Barratt.2022}. The algorithm has to be handcrafted for each problem\cite{Geng.2022}. Second, generalization to higher dimensions is possible \cite{Zaletel.2020,Wall.2021} but not as efficient as for MPS due to entropy scaling\cite{Cirac.2021}.

\paragraph{Global gradient methods} are standard optimization techniques that also apply to TNs. While using an overall global gradient usually is outperformed by renormalization methods, global methods make sense in special cases. In particular, renormalization methods have not been established yet for QTNs. Currently, stochastic gradient approximation methods\cite{Spall.1992} are employed in QML to circumvent the need for costly calculations of total gradients in high dimensional parameter spaces \cite{Grant.2018}.

Global gradients have the downside that the gradient may vanish for random initial conditions in high dimensional parameter spaces. In QML, this is usually referred to as the \emph{barren plateau} phenomenon \cite{McClean.2018} and is similar to the vanishing gradient problem known from classical ML~\cite{Hochreiter.1991}.

The performance of gradient methods can be boosted by considering the special structure of TNs, e.g. with adapted initialization schemes \cite{Barratt.2022}. Introducing locality either on the optimization routine or the loss can also mitigate barren plateaus (see Section \ref{sec:qtn}).

\paragraph{Geometric methods} make use of the network's underlying tensorial geometry. Tools from differential geometry can be used for analyzing the TN on the space of entanglement patterns \cite{Swingle.2012} and optimizing on loss manifolds \cite{Rohwedder.2013}. This kind of optimization performs well on high dimensional parameter spaces, especially in combination with stochastic gradient descent \cite{Novikov.12.05.2016} and auto-differentiation on individual nodes\cite{Luchnikov.2021,Hauru.2021} or whole layers\cite{Geng.2022}.

More advanced geometric methods reuse previous update steps. For this, their gradient vectors have to be transported along the optimization manifold \cite{Cichocki.2016b}. However, they have to be applied in practice yet.

\subsection{Classical Machine Learning with Tensor Networks}\label{ssec:cml}

We already discussed in Section \ref{ssec:generalSt} that TNs are able to approximate high dimensional states within a regular, less complex structure. In ML, such states arise as maps of data features and as weight tensors that connect the data features to the desired result, e.g. a label in classification\cite{EdwinStoudenmire.}. 

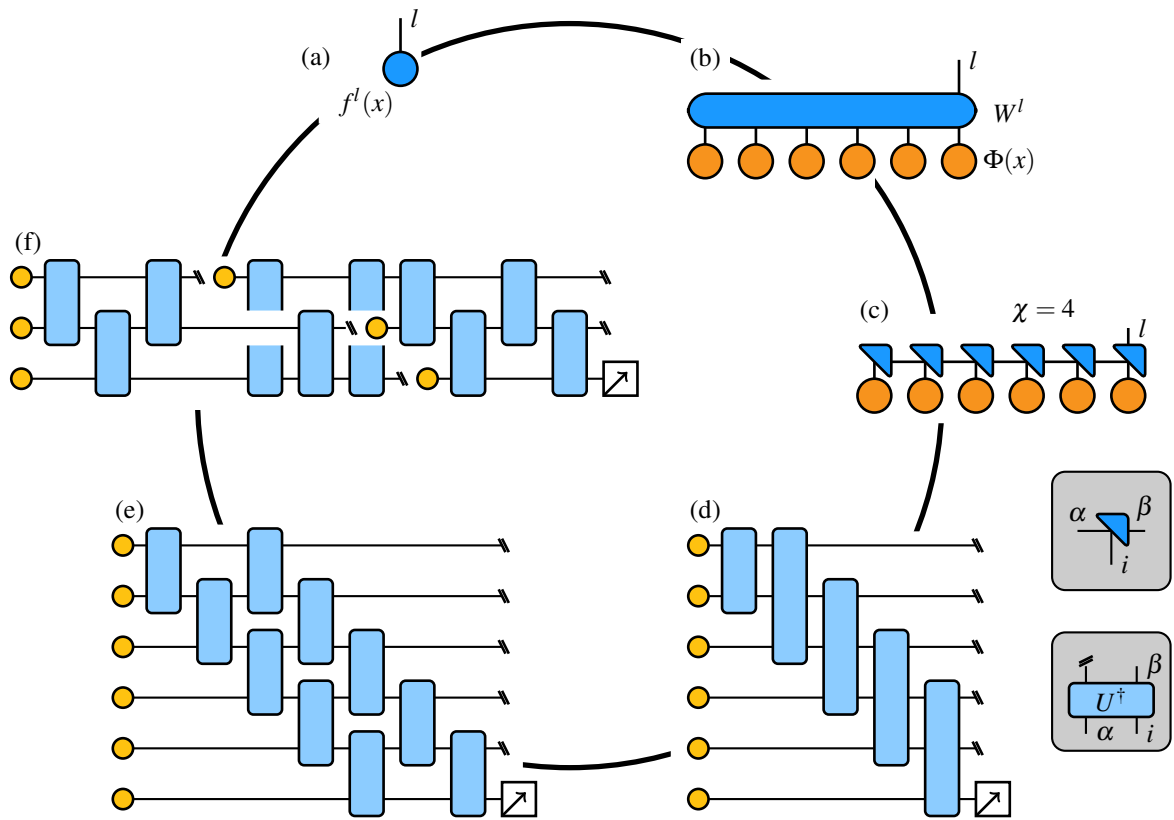
\begin{figure}
    \centering
    \input{Class2QTN.tex}
    \caption[From a classical classifier to an efficient quantum tensor network.]{From a classical classifier to an efficient quantum tensor network. (a) Formally, the task of classification is performed by some function $f_l(x)$ which is an object, that accepts input data $x$ and outputs some label $l$. (b) In machine learning, one realizes the classification function $f_l(x)=\langle W_l|\Phi(x)\rangle$ with a weight tensor $W^l$ with trainable parameters where the (possibly transformed) input $\Phi(x)$ is fed into. The classification function is constructed as an overlap between both tensors. (c) In a tensor network approach, one decomposes the large tensor $W^l$ into a network of smaller tensors, e.g. by restricting the structure to a matrix product state (MPS) layout. In this case, the bond dimension is $\chi=4$. (d) By identifying isometric tensor nodes with unitary quantum gates (grey boxes), the MPS classifier can be mapped to a quantum computer. Higher bond dimensions between the tensor nodes require multi qubit gates. In this case, the resulting circuit needs $\log \chi = 2$ internal qubits and three qubit gates. (e) The multi qubit gates can be expressed by a repetition of the MPS two qubit gate structure. Each additional internal qubit requires another layer of two qubit gates. (f) If the quantum hardware supports resetting qubits during execution, a qubit efficient approach can be implemented reusing discarded qubits. The efficient circuit is a trade-off between qubit number and circuit length.}
    \label{fig:Class2QTN}
\end{figure}

In principle, an ML algorithm seeks to find a function $f_l(x):\mathcal{D}\to\mathcal{S}$ of some datum $x$ within the space of all possible inputs $\mathcal{D}$ that is mapped to a space of possible results $\mathcal{S}$, for instance a set of labels $l$. This function is called the model. Usually, the model is a composition of a data embedding $\Phi(x)$ and a trainable weight tensor $W_l$ connecting the embedded data to the output, as shown in Fig.~\ref{fig:Class2QTN}~(a)-(b). The weight tensor $W_l$ can be approximated as a TN whose output represents the choice of labels (see Fig.~\ref{fig:Class2QTN}, c). We get 

     \begin{equation}
         f_l(x)=W_l\circ\Phi(x)\approx\langle W_{l,TN}|\Phi(x)\rangle
     \end{equation}

where $\langle W^l_{TN}|$ is the TN approximation of the weight tensor. The dimensions of the weight tensor's index $l$ store the probabilities $P^{l_i}$ of the corresponding labels $l_i$ 

\begin{equation}
    W^\chi_{l_i}\circ \Phi_\chi(x)=P_{l_i}(x).
\end{equation}

Multi-class classifications are either done by training a single TN with large outgoing bond dimension or a set of networks with a single outgoing label bond each (one versus all). The data is embedded with a feature map that can transform the data before mapping it to the network\cite{Wall.2021b}. This approach is very similar to encoding maps for QML\cite{Schuld.2021} and can be approximated as TN as well.

A second way of embedding data into a feature space is using a density matrix $|\Phi(x)\rangle\langle\Phi(x)|$ and contracting it with a label dependent weight state $|W_l\rangle$. In this construction, the bond dimension $\chi$ is given directly by the non-vanishing eigenvalues of the covariance matrix\cite{Wall.2021} and the decision function is realized as the maximum overlap
	
	\begin{equation}
	    f_l(x)=\text{argmax}_l \langle W_l|\Phi(x)\rangle\langle\Phi(x)|W_l\rangle.
	\end{equation}

This construction has the advantage of being able to process incomplete data by contracting over missing bonds and can represent specific probability distributions based on the data sets\cite{Wall.2021b}. 

Building generative TN models is also straightforward. The goal of a generative ML model is to learn the distribution of its training data and to generate additional samples from this distribution. The simplest possibility is to use the dual of a trained classifier or a regressor obtained by adjoining all tensor nodes within the network. 

Due to their quantum inspired construction, TNs have the issue of not being able to copy information within their structure. This means, that information cannot be distributed to different branches of a TN in a way a neural network uses information to activate its neurons for example. If for instance an operation in image analysis needs to use the value of adjacent pixels, one has to pass the same data into several input nodes by using overlapping observation windows\cite{Glasser.15.06.2018}. However, this approach does not allow copying connected tensors to different locations.

Often, it makes sense to combine different layouts to use advantages of both. As an example, hierarchical layouts coarse grain the data and grid layers can be used to efficiently combine the information from different branches of the hierarchical TN \cite{Reyes.22.01.2020}. The hierarchical part can be optimized with unsupervised ML methods where the ideal weight tensor is derived from the data covariance matrix\cite{Stoudenmire.2018}. It is even possible to add a TN layer to a neural network architecture e.g. for complexity reduction in the input layer with MPS\cite{Chen.2018}, MERA convolutional layers \cite{Kong.} or approximating a fully connected layer \cite{Novikov.2015}.

TN architectures are closely related to neural networks. Restricted\cite{Chen.2018,Glasser.2019} and deep\cite{Li.2021} Boltzmann machines can be mapped to a two dimensional TN consisting of MPS and Matrix Product Operators (MPO), an operator valued version of MPS. Boltzmann machines therefore may be simulated using an MPS which allows for adjusting accuracy and execution time via the bond dimension allowing for a compression of neural network representations\cite{Cichocki.2016b}.

The map between both architectures has been exploited in both ways to compare specific network layouts. On the one hand, node numbers in an MPS representation of a Boltzmann machine will scale exponentially with the number of neurons\cite{Collura.2021} and recurrent neural networks can simulate MPS with reduced computational effort for certain cases\cite{Wu.24.06.2022}. On the other hand, hierarchical TNs efficiently implement convolutional or recurrent neural networks\cite{Levine.2019}. 

\paragraph{Applications in ML} can been found for a wide variety of tasks. In image analysis, TN based ML models are used for classification\cite{Wall.2021,Araz.2021,Felser.2021}, compression\cite{Selvan.13.11.2020} or feature extraction \cite{Kong.,Liu.2021}. TN based regressors have been successfully applied to nonlinear system identification\cite{Batselier.2022} where the task is to generate a model of a nonlinear system from its behaviour. 

Generative TN structures have been employed in anomaly detection\cite{Wang.03.06.2020} and as classifiers when reversing the generative TN structure\cite{Sun.2020}. A TN can be used to learn a probability distribution from data and a simulated "measurement" of the TN state will generate a new instance from the distribution\cite{Han.2018,Cheng.2019}. Generative TNs have also been applied to unsupervised feature identification in images\cite{Bai.2022}.

\section{Quantum Tensor Network Machine Learning} \label{sec:qtn}
\subsection{Mapping to Quantum Circuits}\label{ssec:generalSt}
	
	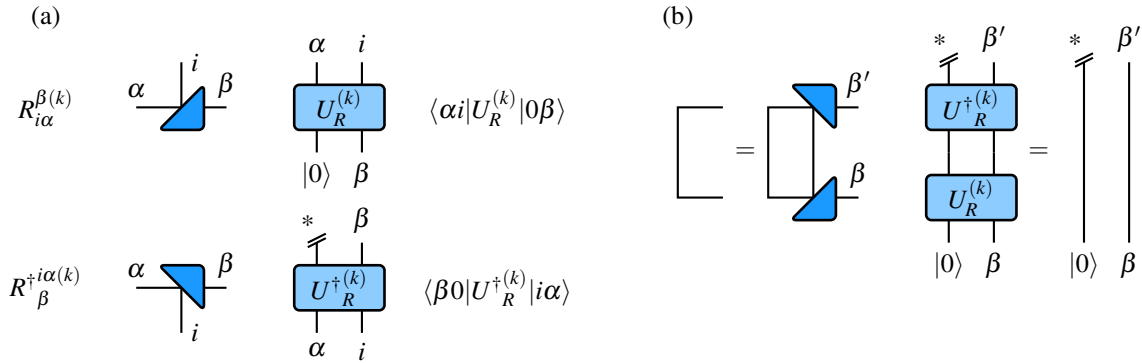
\begin{figure}
	    \centering
    \input{TN2QTN}

	\caption[Mapping from isometric tensors to unitary quantum gates]{Mapping from isometric tensors to unitary quantum gates. 
	(a) Whether a bond $\alpha,\beta$ is mapped to an incoming or outgoing qubit bundle, depends on whether the tensor is given in right or left isometric form. The free bonds $i$ are represented by outgoing qubits. Adjoining a tensor flips its directions. Qubit preservation is taken into account by adding additional ancilla qubits or discarding left over ones. 	
	(b) Mapping the normalization condition for isometric tensors illustrates that discarding qubits actually has to fulfil a condition: For an exact representation of the classical network, discarded qubits have to be post-selected to $|0\rangle $ to be the dual of the ancilla state.}
	    \label{fig:TNtoQTN}
	\end{figure}

The quantum-inspired construction of TNs makes it straightforward to translate the concept to quantum computations. Tensor nodes are realized by multi-qubit gates with incoming and outgoing qubits carrying the bonds of the node. 

The procedure for mapping the classical TN to a QTN is shown in Fig. \ref{fig:Class2QTN} (c)-(d). Quantum gates are unitary therefore the corresponding TN has to be in canonical form with at least isometric nodes\cite{Liu.2019} (see Section \ref{ssec:layout}). Fig. \ref{fig:TNtoQTN} (a) shows how isometric tensors are mapped to gates. 
The bond dimension $\chi$ is determined by the number of qubits $n$ transferred between connected gates, i.e., $\chi=2^n$. These qubits are called internal or virtual qubits. 
The qubits carrying the free (or physical) bonds are either forward or backwards directed, depending on whether the node has a vector or dual valued index. To preserve the number of qubits, the sum of all incoming qubits (free and internal) must equal the sum of the outgoing qubits at each gate. Therefore, one prepares necessary additional incoming qubits in a dummy state $|0\rangle$ or discards left over outgoing qubits. 

Discarded qubits usually are carried on unobserved, but a direct correspondence to classical TNs requires post-selection to a reference state $\langle 0|$ on these qubits\cite{Wall.2022}. From the normalization condition in Fig.~\ref{fig:TNtoQTN} (b)

\begin{equation}
    	\delta_{\beta'}^\beta=R^{\beta(k)}_{i\alpha}{R^\dag}^{i\alpha(k)}_{\beta'}\quad\quad
    	\leftrightarrow\quad\quad\langle\beta' *| {U^\dag}^{(k)}_RU^{(k)}_R | 0\beta \rangle = \langle *|0\rangle \langle\beta'|\beta\rangle
\end{equation}

it follows that a post-selection measurement on $\langle*|$ is the counterpart of the ancilla $|0\rangle$ initialization. This is caused by the fact that an isometry is mapped to a unitary and classical dimensional reduction or information loss has to be accounted for. Instead, one also can perform an uncomputation operation for each gate used\cite{Grant.2018}. For a network fully optimized on the quantum machine, the post selection requirement can be released which allows for hybrid methods \cite{Wall.2022,Dborin.2022} or efficient layouts where discarded qubits can be reset and reused \cite{Huggins.2019}. 

\begin{figure}
	    \centering
	    \input{MPSlayouts}

    \caption[Simple tensor networks and their quantum counterparts]{Simple tensor networks and their quantum counterparts. Single qubit unitary gates are omitted as they can be absorbed into an adjacent two qubit unitary. Each bond may be realized by one or more qubits.
    (a) shows a matrix product state (MPS) in site canonical form and its quantum implementation. Choosing a central gauge halves the circuit depth compared to the left canonical MPS from Fig. \ref{fig:Class2QTN}. However, the central node is not isometric in general and can only be mapped to the unitary quantum gate approximately. If the quantum computer supports resetting qubits during execution, a qubit efficient approach can be implemented reusing discarded qubits with constant qubit number. (b) A tree tensor network offers higher entanglement than a matrix product state, but its qubit efficient quantum representation will need a total of $\log n$ qubits for $n$ inputs.
}
	    \label{fig:MPSLayouts}
	\end{figure}
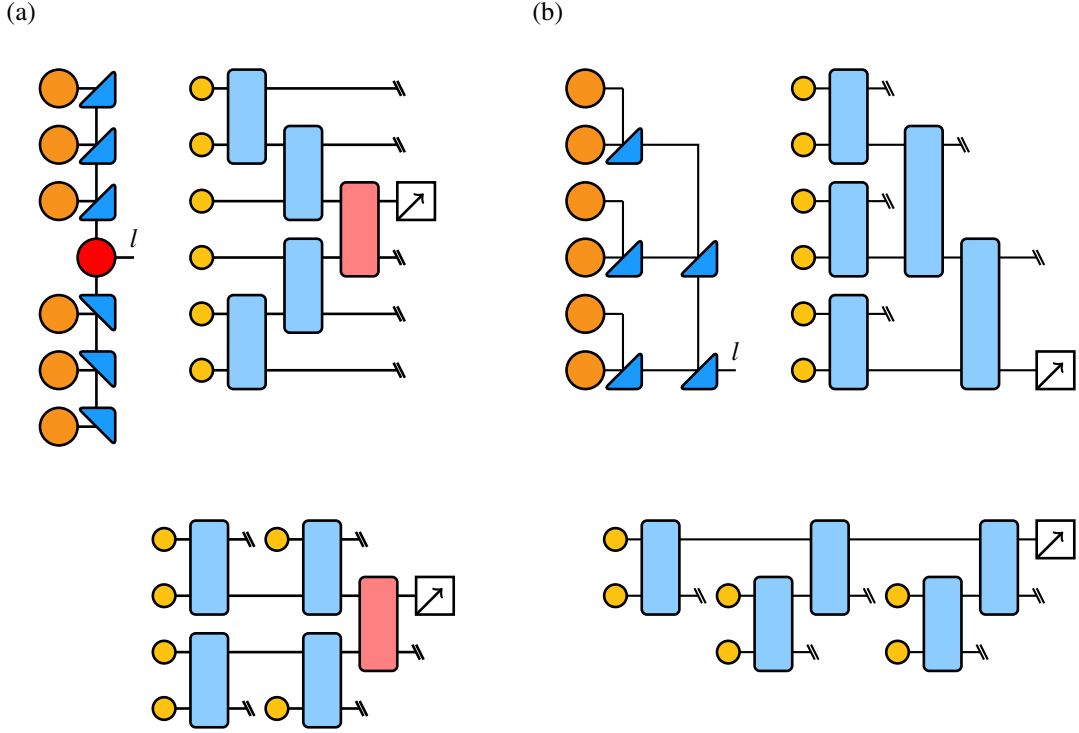

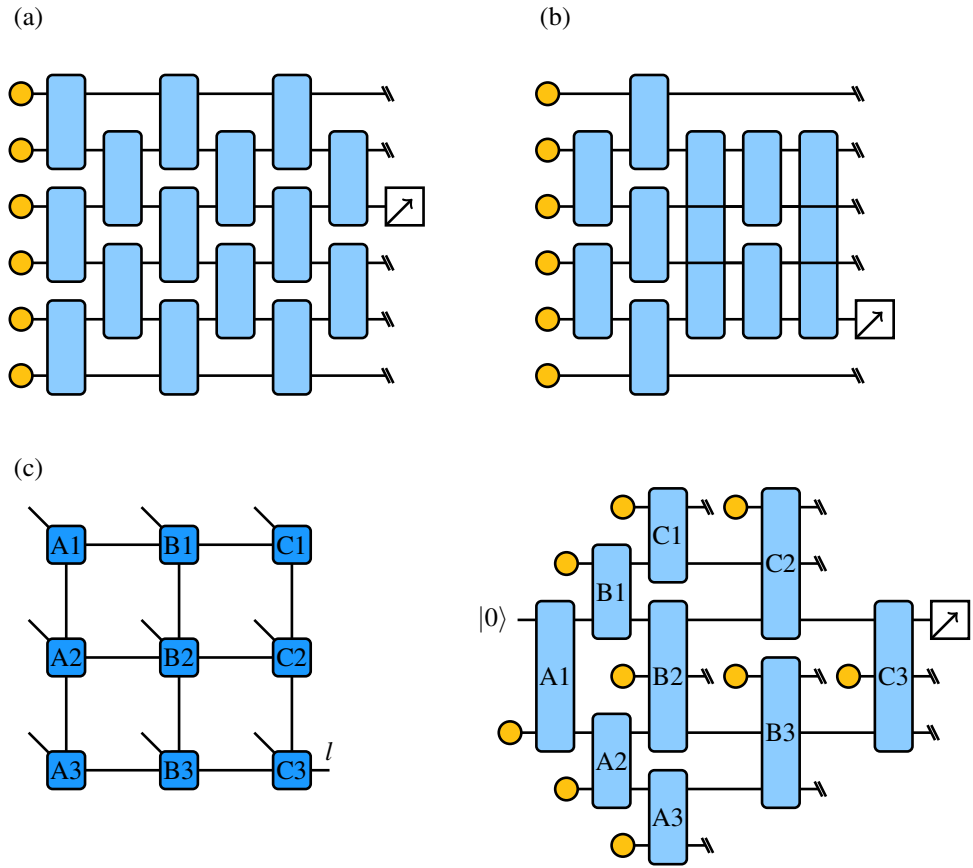
\begin{figure}
	    \centering
        \input{TNLayouts}

	    \caption[Higher dimensional quantum tensor network structures]{Higher dimensional quantum tensor network structures. (a) The brickwall architecture offers a higher amount of entanglement than matrix product states (MPS) and can be seen as a derivative of hexagonal projected entangled pair states. A brickwall allows for the representation of every MPS gauge up to a bond dimension given by the depth of the circuit. (b) The multiscale renormalization ansatz (MERA) quantum network requires gates between qubits further apart which may be realized by introducing swap gates in between on current hardware. Both brickwall and MERA do not allow for qubit efficient implementations. (c) The quantum circuit of pair entangled product states (PEPS) heavily depends on the order in which the PEPS nodes are evaluated. The realization will feature coupled staircase structures similar to MPS. Here, a qubit efficient approach scales linearily with the length of the diagonal.}
	    \label{fig:TNLayouts}
	\end{figure}

Using this recipe, one can map the TN layouts known from Section~\ref{ssec:layout} to quantum circuits. Fig. \ref{fig:MPSLayouts} shows a central gauge MPS and a TTN. 
More advanced networks like a brickwall, MERA and a square PEPS are shown in Fig. \ref{fig:TNLayouts}. Mapping these networks to a quantum computer gets more and more involved with growing bond dimension and requires larger circuits with high connectivity (or many swap gates) between the qubits.

\subsection{Efficiently implementing Quantum Tensor Networks}

Large circuits, multi-qubit gates and gates not using the standard gate set are hard to implement on near-term noisy intermediate scale quantum (NISQ) computers. To reduce circuit complexity, several approaches exist. A major step in bringing TNs to quantum computers was the development of a breakdown method for multi-qubit nodes to two-qubit unitaries with high fidelity~\cite{Ran.2020, Lin.2021}, which can be implemented on NISQ devices efficiently. This approach has been based on a classical procedure for photonic qudits\cite{Schon.2005}. The approach is shown for an MPS in Fig.~\ref{fig:Class2QTN}~(d-e): This MPS has left canonical gauge and therefore has a quantum gate equivalent looking like a staircaise of multi-qubit gates. The size of the gates is given by the number of internal qubits $n=\log\chi$ that have to be passed on to the next gate (d). The three qubit gates in this example may be replaced by two layers of two qubit gates which provide the same connectivity between adjacent incoming free qubits (e). Each additional internal qubit would add another layer of two-qubit gates to the circuit. 

The most general ansatz for the gates within a tensor node is a full unitary gate\cite{Dilip.24.04.2022,Huggins.2019}. Representing these gates with simple gates available on a NISQ device however results in long circuits that are prone to noise. Therefore, simplified ansaetze for the two-qubit gates are commonly used\cite{Guala.2022, Fastovets.01.10.201805.10.2018,Araz.21.02.2022, Uvarov.2020,Dborin.2022}. For quantum input, the performance of these simplified ansaetze can yield comparable maximum performance to a general unitary ansatz, but they seem to be harder to train. For classical data, the performance was much lower using simple nodes. This holds for both grid \cite{Lazzarin.2022} and hierarchical layouts\cite{Grant.2018}.

To reduce the total qubit count, the structure of many TNs allows for an efficient reordering of its blocks, such that discarded qubits can be reset and reused for the input of new information (see Fig.~\ref{fig:Class2QTN}~f). A qubit efficient MPS only requires a constant amount of qubits determined by the dimension of the inputs and the desired bond dimension. For a TTN, the qubit number scales logarithmically with the size of the input.

Using the qubit efficient approach may not have an effect on the optimized model parameters because the circuit is trained to carry on the label information and the 'no signalling' principle therefore forbids an influence of these discarded qubits on the result\cite{Huggins.2019}. Instead of simply resetting, the information in the discarded qubits can be used for quantum error correction within the nodes\cite{Cong.2019} which improves performance on NISQ devices. In general, the influence of the qubit efficient procedure e.g. on trainability is still not clear. When combined with a local loss however, no barren plateaus arise in the error correcting ansatz\cite{Pesah.2021}.

Combining these simplifications reduces both qubit number and gate complexity \cite{Ran.2020}. The overall circuit depth is harder to reduce. Choosing a central gauge for MPS at least halves circuit depth compared to left or right gauges\cite{Dborin.2022} (see~Fig.\ref{fig:MPSLayouts}).

\subsection{Variational Machine Learning with Quantum Tensor Networks}\label{ssec:vqc}

Recently, a wide variety of ML architectures employing variational quantum circuits (VQC) have been developed. A VQC is a quantum circuit whose gates have tunable parameters. General unitaries can be constructed from a combination of rotation and entangling gates like the CNOT gate. A common architecture for QML is a layered VQC. Here, the circuit consists of encoding blocks that map the data to the circuit and parametrized variational blocks which entangle the qubits. 
To increase the expressivity of the quantum circuit, these blocks can be repeated before the measurement \cite{Schuld.2021}. 

A QTN with tunable gates is also a variety of VQC with an internal TN layout. The structure of TNs provides several advantages for QML. First of all, insights from the available theory on classical TNs also apply to their quantum counterparts. Due to the direct correspondence, data and models from classical TNs can be translated to QTNs and vice versa. This can be used to better initialize quantum models (see Section~\ref{ssec:hybrid}). Furthermore, the choice of a specific TN layout allows for the introduction of inductive bias, e.g. knowledge about the type of data and therefore the construction of a QML structure that will fit the data well. Finally unlike for general VQC algorithms, the space of possible weights in TN based ML can be adjusted easily by varying the bond dimension. This allows for tuning the expressivity of the circuit to mitigate under- and overfitting \cite{Huggins.2019}. It is not clear yet how the expressivity of a QTN scales or compares to layered VQC approaches but both architectures can be mapped onto each other\cite{Du.2020}.

Until now, the development of QTN ML approaches focuses on supervised classificators and generators. Supervised learning with QTNs works similar to the classical approach shown in Section~\ref{ssec:cml}. Examples for QML circuits based on different layouts are shown in Figures \ref{fig:MPSLayouts} and \ref{fig:TNLayouts}. The data is mapped to the quantum computer using some feature map $\Phi(x)$ which is shown as orange dots in the images. The feature map may be a tensor network itself (see Section \ref{sec:Encoding}). The weight tensor $W_l$ is represented as the blue quantum tensor network. In the end, a measurement on the remaining qubits yields a result, e.g. a classification. If a multiclass output is needed, introducing an exit node (see Fig.~\ref{fig:EncStrat}~b) will improve the fraction of correct classifications\cite{Dilip.24.04.2022}. 

Generative TNs can be realized by reversing the TN structure. The inputs of the generative network are given by some reference computational basis state which are entangled by the TN (see Fig.~\ref{fig:EncStrat}~a). These generative networks can be trained either by sampling the generative QTN and comparing the results to a given training set\cite{Huggins.2019} or by training a classifier and adjoining every gate as noted in Section~\ref{ssec:cml}.

Some studies already include an investigation of the influence of noise on the QTN circuit. Numerical results indicate, that low level noise is not a problem for classification \cite{Huggins.2019,Grant.2018}. It even may be used to enhance the performance of the algorithm by adding ancilla qubits initialized with noise to the circuit. This effectively generates a probabilistic model which is easier to train. However, if the noise is too high this also leads to decoherence rendering the circuit unfunctional\cite{Liao.02.09.2022}.

Optimizing the parameters of these QTNs relies on some variety of global gradient descent for the majority of literature. Geometric \cite{Wall.2022} or genetic methods \cite{Chen.2022} are used only rarely. Renormalization methods like for classical TNs have not been adapted to QTNs but may be employed in hybrid methods \cite{Lubasch.2020}. Some proposals even consider employing TNs for optimizing parameters or hyperparameters of QML algorithms\cite{Sagingalieva.10.05.2022}.
For specific implementations, first evidence exists that the locality of TNs can overcome barren plateaus \cite{Zhang.12.11.2020, Qi.08.06.2022}. Especially the use of local loss functions, which can be implemented using local Hamiltonians, provides a favourable loss landscape without gradients vanishing exponentially fast\cite{Pesah.2021,Liu.2022}. The same approach may also reduce the amount of training data needed\cite{Araz.21.02.2022}. 

\subsection{Hybrid training}\label{ssec:hybrid}

Hybrid QTN architectures combine quantum and classical elements to use the advantages of both worlds. Compared to NISQ devices, classical computers are able to perform computations on far larger datasets and their use is very cheap. The quantum part of the algorithm may introduce some qualitative quantum advantage like higher maximum performance or generalization of the model. At the moment, two hybrid strategies make use of these characteristics. First, the classical reduction of the input data's dimensionality with pre-processing like PCA\cite{Dborin.2022}, auto-encoders or TN based encodings discussed in Section \ref{sec:Encoding}. If the classical part is trainable, it may be optimized together with the subsequent QTN. Second, the direct maps between TNs and their quantum counterparts allow for classical pre-training of the quantum model's initial values. Even when more powerful quantum computers are available, the execution of quantum circuits will still be expensive and pre-training methods to reduce the number of quantum circuit executions will stay relevant. In this section, we will focus on hybrid pre-training methods. 

As discussed in Section~\ref{ssec:generalSt}, a TN in canonical form can be mapped exactly to a quantum computer. This allows to train a coarse classical TN model which can be refined and expanded after mapping it to a quantum computer. Any standard QTN layout may be prepared with classically prepared initial conditions\cite{Wall.2021,Wall.2021b} and for providing efficient initial values, no post-selection on the quantum computer is required \cite{Wall.2022}. Using these initial values for the QTN's parameters makes the training of larger quantum circuits far more efficient in comparison with random or identity initialization schemes. The main benefit is that the initial training phase, where the gradients decrease exponentially with the qubit number already has been performed classically and therefore the training on the quantum device starts in a favourable spot of the parameter space \cite{Dborin.2022}.

    \begin{figure}
	\centering
	\input{QTNWall}

        \caption[Hybrid training methods for quantum tensor networks.]{Hybrid training methods for quantum tensor networks. A pre-trained classical tensor network provides suitable initial values for further optimization on a quantum computer. The direct approach may be refined to make training easier or to have access to a larger part of the multi-qubit Hilbert space. Method (a)\cite{Dborin.2022} maps a classically optimized MPS to the diagonal gates $U_c$ of a brickwall ansatz; all off-diagonal gates are initialized as identities. Then a second optimization step on the quantum computer is conducted. While the optimized diagonal gates $U_o$ have to be full unitary gates to enable the transfer from the classical network, the off-diagonal gates $W$ can use a simpler ansatz with fewer parameters. Approach (b)\cite{Wall.2022} maps two classical MPS to the quantum circuit. A homogeneous MPS ($N_b$-times $U_G$) truncated to an appropriate boundary condition $U_R$ prepares an initial state. In the second MPS, the nodes $U_D^{(i)}$ upload and process the $N_x$ elements of the datum $x$. Finally, the classification is performed on the exit node $U_C$. Method (a) is shown as a generator, method (b) as an efficient classifier.
        }
	\label{fig:QTNWall}
    \end{figure}
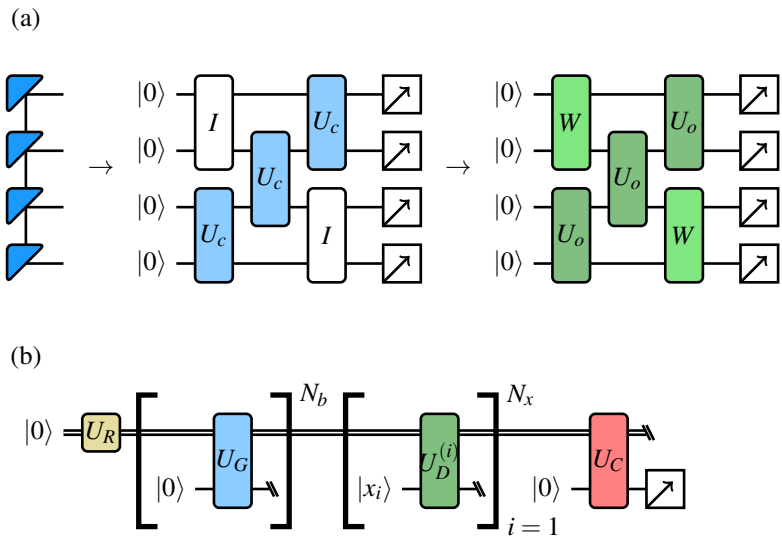

Modifications to the basic procedure of classically pre-training a QTN have been developed to lower the requirements on the classical preparation and to make the quantum part easier to train. For training a brickwall layout (see Fig. \ref{fig:TNLayouts}~a), it is sufficient to prepare an initial MPS state that is embedded within the brickwall, e.g. the diagonal, and the remaining gates start as identity gates\cite{Dborin.2022} as shown in the centre panel of Fig.~\ref{fig:QTNWall}~(a). To make quantum training easier, one does not have to use full unitary gates $U$ on the whole circuit, but can restrict the off-diagonal gates to some simple ansatz $W$ as shown in the right panel of this figure. This approach can be seen as a quantum version of the copy node initialization for classical TNs, where most of the tensor nodes are initialized with identity tensors \cite{Barratt.2022}.

Another modification considers preparing a prior distribution within the feature space before uploading the data. This reduces the bond dimensions and gate complexity needed \cite{Wall.2022}. Fig.~\ref{fig:QTNWall}~(b) shows the approach for an efficient MPS classifier. At the beginning of the circuit, a homogeneous MPS $U_G^{N_b}$ of length $N_b$ prepares the prior distribution. Setting a trainable boundary condition $U_R$ reduces the number of nodes the MPS needs to represent an effective prior. The second part $U_D^{(i)}$ is a standard efficient MPS similar to Fig. \ref{fig:Class2QTN}~(f), where the $N_x$ data features are introduced into the QML and an output node $U_C$ prepares the classification result in the end. The circuit technically can be optimized without classical pre-training. But for higher bond dimensions, this construction is far easier to train having initial values obtained with classical TN-specific methods like DMRG \cite{Wall.2022}.

\subsection{Case Studies and Implementations}\label{ssec:TN4QML}

The application of QTN ML methods has been limited to demonstrative feasibility studies up to now. Most authors focus on classification tasks for image recognition either with binary classes\cite{Araz.21.02.2022} or multiclass setups\cite{Dilip.24.04.2022}. One implementation of binary image classification\cite{Huggins.2019} has been performed on real photonic hardware\cite{Wang.2021}. Other uses are classifications on parameterized classical data\cite{Bhatia.04.05.2019} and on quantum simulation results \cite{Uvarov.2020, Lazzarin.2022}. Besides the proof of concept, these studies demonstrate that QTN approaches already can process relatively high dimensional input data like grayscale images of up to $37^2$ pixels. They show that QTNs can achieve accuracies for the classification of both classical and quantum datasets in the range of $0.85$ to $0.95$ with only a small amount of parameters and internal qubits. 

The application of QTNs for a regression of continuous properties has not been discussed widely yet. One proposition for this application is to approximate eigenvectors of unitary matrices\cite{Kardashin.2021}, but finding the right bond dimension is crucial to find an approximate state having sufficient overlap with the real eigenvector without using huge circuits.

QTN generators have been implemented by various authors to provide quantum state samples from learned distributions\cite{Huggins.2019, Wall.2021b, Dborin.2022} as a feasibility study.

Most case studies that use publicly avaiable frameworks rely on the qiskit \cite{Qiskit}, as it supports resets in the middle of an execution, which are necessary for efficient TNs. For ML, Qiskit is compatible with the pytorch framework. Cirq \cite{CirqDevelopers.2022} also provides a reset functionality and integrates with the tensor flow ML suite. Pennylane\cite{Bergholm.12.11.2018}, which focuses on QML applications, currently cannot implement mid-circuit measurements for efficient QTNs, but provides methods for both basic MPS and TTN based quantum classifiers. The implementation allows for varying virtual qubit bonds and connects the quantum circuits to most common ML frameworks in Python.

\section{Tensor Networks for Data Encoding}\label{sec:Encoding}

For the performance of data driven quantum algorithms and QML algorithms in particular, encoding of data plays a crucial role. Current quantum computing hardware provides neither a sufficient amount of qubits nor gate depth to encode high dimensional data sets in a straightforward fashion. However, this does not necessarily mean that the problem size to be tackled with current quantum algorithms has to be small. Instead, one relies on classical and quantum pre-processing steps that reduce the data to its essential features. 

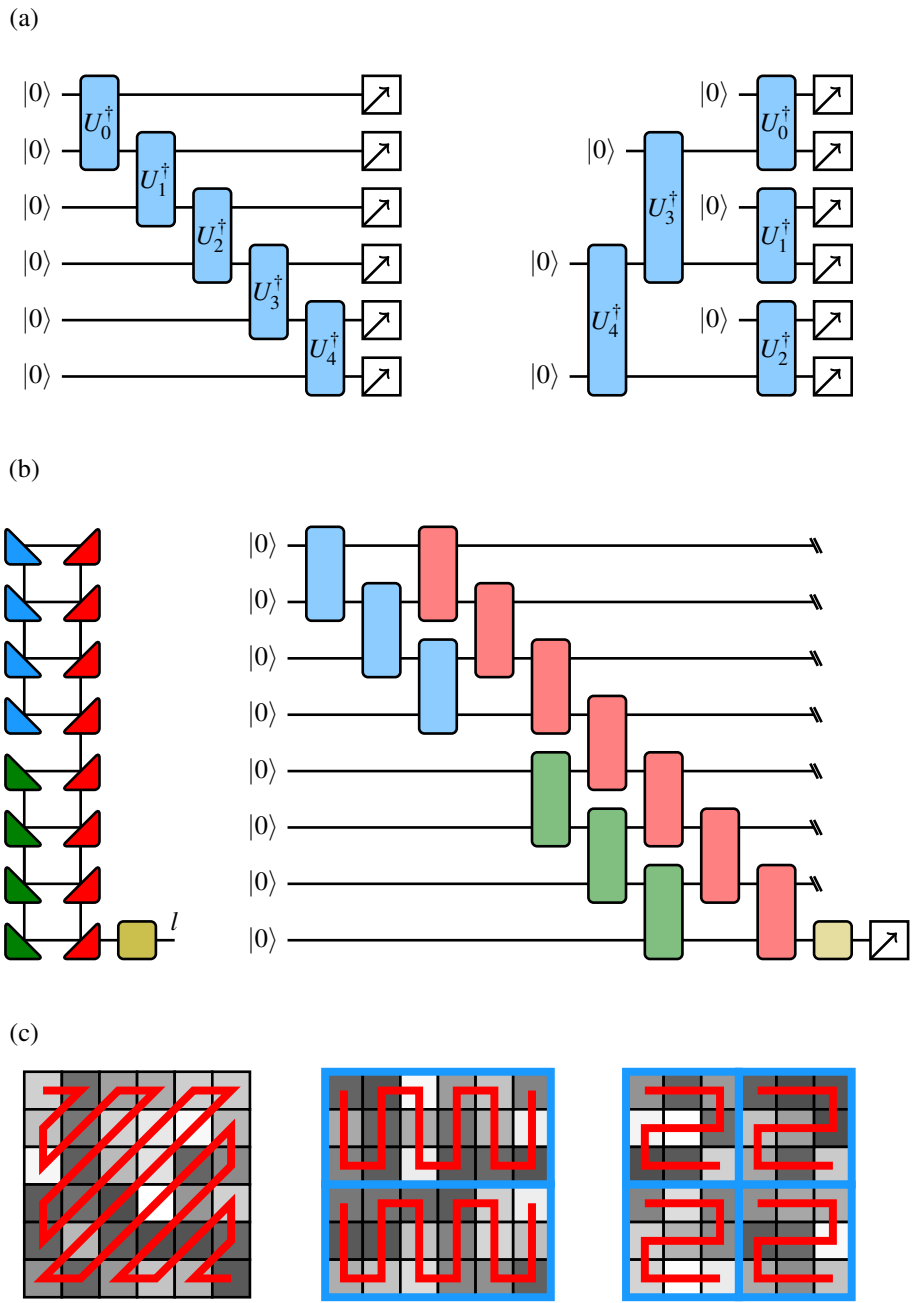
\begin{figure}
	    \centering
	\input{EncStrat}
    \caption[Encoding strategies for machine learning using tensor networks]{Encoding strategies for machine learning using tensor networks. (a) Encoding classical or compressing quantum data using matrix product states (MPS, left) or a tree tensor network (TTN, right). Each gate $U^\dag_i$ is a direct mapping from an isometric node of a classical tensor network. A generative quantum tensor network has the same structure as an input state, but with one or more $|0\rangle$ input qubits replaced by a label or noise encoded input. (b) Data may be encoded in several independent MPS (blue, green) and fed into the circuit to reduce information loss. A quantum machine learning algorithm can make use of the same MPS structure (red) and directly connect to the encoding MPS. An additional output gate (yellow) improves classification accuracy for multi-class tasks. (c) To encode two-dimensional data like images into MPS, one needs to choose a one dimensional path either at the cost of losing parts of the information or introducing high bond dimensions. Cutting the area into patches improves the encoding result as this reduces the maximum distances on the MPS between neighbouring sites in the original data. 
    }
	    \label{fig:EncStrat}
	\end{figure}

By adjusting the bond dimension, TNs provide a direct way of compressing data both lossless by discarding dimensions with singular values equal to zero and approximate by setting upper bounds on the bond dimension. The input QTN state can be prepared in at least five different ways. First, by maximizing the overlap between a classical representation of a quantum state and a TN. Second, by encoding classical data in a TN and reducing its bond dimensions by tensor decompositions. In both approaches, the network will then be mapped to quantum circuits as shown in Fig.~\ref{fig:EncStrat}~a for MPS and TTN. Efficient mapping methods are also available for PEPS\cite{Schwarz.2012}. A third classical method is training a TN to compress the data into a latent representation with fixed bond dimension and encode the latent vectors using some direct strategy. On a quantum computer, one can directly maximize the overlap between an existing quantum state and a QTN. This requires preparing the reference quantum state multiple times until convergence which may be very costly. Finally, one can train a generative network to output a state from some distribution (see Section~\ref{ssec:vqc}). Therefore we focus on classical pre-processing in this section. 

Encoding data in TN based quantum layouts promises some benefits over classical encoding. Especially, the access to a large state space even when using few qubits could boost efficiency of information storage. For example, the number of parameters needed to represent certain time evolutions of quantum states is exponentially reduced with QTNs compared to classical ones\cite{Lin.2021}.

Depending on the type of data, different layouts of TNs provide the most efficient storage because the scaling of their mutual information has to match the scaling behaviour of entanglement in the TN. As discussed in section~\ref{fig:MPSLayouts}, 
MPS suit 1D-Data like time series and logarithmic TTNs or 2D TNs like MERA or PEPS are better suited for images depending on the amount of local correlation. For text, the information scales even steeper\cite{Lu.11.03.2021}, which requires 3D-PEPS or high dimensional MERA variants which have not been implemented on a quantum computer yet. Exploiting symmetries reduces the need for complexity within the structure, e.g. by using wavelet transform techniques in images \cite{McCord.04.03.2022}.  

Nevertheless, MPS and TTN can be implemented and optimized easily and still provide an improvement over direct encoding methods. They are therefore widely used in encoding for QML. The performance of MPS and TTN can be improved by combining them with other methods. When encoding images, one can split the whole image into patches and encode each patch into an MPS (see Fig.~\ref{fig:EncStrat}~c) which will catch local entanglement better but requires more storage. For a fixed bond dimension, the number of qubits is proportional to the number of patches encoded. The pixels in each patch are addressed by a method that is known as flexible representation of quantum images (FRQI)\cite{Le.2011}. The method was developed as a classical compression method \cite{Latorre.2005} and has recently been transferred to quantum computers\cite{Chen.2018,Ran.2020}. Patchwise MPS encoding can be easily combined with MPS QML methods (see Fig.~\ref{fig:EncStrat}~b)\cite{Dilip.24.04.2022}.

Trainable TN encoding using a latent space representation from the outgoing bond dimensions usually is optimized together with the parameters of the QML circuit \cite{Araz.21.02.2022, Chen.2021}. A theoretical study on the error performance of function regression models finds upper bounds when certain continuity requirements on the loss and the network are met\cite{Qi.08.06.2022}. Particularly, they find that the optimization error connected to barren plateaus will be negligible if the loss on the TN parameters is Lipschitz and satisfies a Polyak-Lojasiewicz condition.
However, they do not develop a method to set up a TN that actually fulfils these conditions. Trainable encoding can be improved by a patchwise approach, too. Applying trainable MPS approximators on small regions of the image yields a linear model of the image where the spatial information is stored in the feature space\cite{Selvan.13.11.2020}. Due to the independence of the various layers, this method also could be realized with a hybrid circuit, where the initial layers are classical and the final layers are quantum.
	
TN encoding pairs well with TN based ML but it is applicable to any other QML approach. For layered VQC approaches, first results imply that TN pre-processing trained together with the VQC classifier performs better than regular PCA on image data\cite{Chen.2021} and can be used as an estimator for the Q-value function of a reinforcement learning ansatz\cite{Chen.2022}.

TN encoding is not only relevant for QML, but can be used to provide states for any other quantum application that requires complex input. For example, overlaps of QTN generated basis functions can be used to approximate non-linear functions\cite{Lubasch.2020}. This approach may reduce the number of grid points needed in quantum simulations with nonlinear PDEs as couplings compared to the classical approach.

\section{Conclusion}

TNs have proven themselves useful for storing and processing quantum states as well as for classical ML applications. Combining both aspects makes them a suitable tool for QML as well. We have seen in Sections \ref{sec:qtn} and \ref{sec:Encoding}, that TNs can be employed for various tasks within the QML pipeline, from pre-processing and encoding to the variational part and the optimizers\cite{Sagingalieva.10.05.2022}. They have a very flexible representation as they allow for both pure quantum algorithms and classical-quantum hybrids while a wide range of optimization methods can be applied. 

Bringing TNs to a quantum computer has advantages considering architecture design. The representation of a quantum state with TNs on a quantum computer reduces the necessary amount of qubits compared to other encoding methods \cite{Barratt.2021}. Thereby TNs provide an efficient way of mapping classical data to quantum applications. The tensors of a QTN do not have to be contracted costly as on classical hardware since the contraction happens as part of the execution of the quantum circuit. When using hybrid approaches, TNs allow for a seamless connection between classical and quantum methods enabling pre-training and gradual tuning of the border between both systems -- which will become important when the power of NISQ devices scales up significantly. Having the possibility of choosing a qubit efficient implementation is also a very important feature although its effects on trainability are not yet fully understood and require further investigation.

In comparison to classical TNs, QTNs are expected to provide several benefits for the algorithms themselves. As quantum algorithms naturally implement entanglement, QTNs will have access to a  Hilbert space that grows exponentially with the number of qubits. This enlarges storage capacity and the available parameter space for QML algorithms. While classical TNs are able to represent only low-entangled, low-complexity states, QTN have access also to low-complexity states that can be generated by Hamiltonian time evolution \cite{Lin.2021} independent of the amount of entanglement. However, this may need very large circuits depths. Additionally, QTNs provide a natural way of using complex numbers instead of real ones which reduces the number of parameters necessary greatly in certain architectures \cite{Glasser.2019}. It is yet unclear this is a general advantage of QTNs.

Regardless of that, QTNs seem to be easier to train than other QML methods. For example, utilizing local optimization routines that make use of the localized TN structure can help to overcome problems like barren plateaus and reduce the amount of training data needed. However, these results have been obtained using specific implementations, some combined with special features like error correction. The results are therefore not generalizable to all QTN layouts yet. Choosing a layout that fits the data structure well also can reduce the need for large general circuits that are hard to train due to their large amount of parameters. 

The mentioned actual and possible benefits come with downsides compared to classical networks. Gates are directed and reshaping the network cannot be performed in a straightforward way. The usual difficulties with quantum computations like encoding classical data and the need to perform non-reversible measurements to obtain a result still apply. Moreover, compared to more general QML methods like layered VQCs, the strict structure of a TN layout may render it an architecture which cannot be applied on general problems but has to be handcrafted each time. Therefore it is unclear at the moment, whether the benefits of TNs really can be translated to a relevant quantum advantage outside the lab.

Although QTNs have the potential to be a successful framework for QML, their development has just begun and further research is needed in many directions. Modifications to the basic layouts like variable bond dimensions which can be used to reduce computational costs have not been adapted to QTNs yet. In particular, a quantum version of TN-specific local optimization methods is interesting for building algorithms that can be trained more easily. 

However, most important is a more fundamental insight into the capabilities of QTNs -- especially in comparison with classical or other VQC based methods. This includes methods to assess ML performance theoretically on the layout level and not just for specific implementations. Having general measures e.g. for expressivity or trainability would enable us to identify the range of application where it makes sense to use QTN architectures and concentrate future development on these areas.

\bibliography{main}

\section*{Acknowledgements}
The authors thank 
Bogusz Bujnowski, 
Lautaro Hickmann, 
Markus Lange 
and Pia Siegl for our discussions on classical TNs, QTNs and ML and their very helpful remarks on this review.

\section*{Author contributions}
The main contribution was done by Hans-Martin Rieser. 

\section*{Competing interests}
The authors declare no competing interests.

\end{document}

%% file: tikzstuff.tex
\definecolor{lblue}{rgb}{.1,.6,1}
\colorlet{dyell}{yellow!75!black}
\colorlet{dgreen}{green!50!black}
\colorlet{orng}{yellow!50!red}

\newcommand{\TNode}[4]{ 
	\coordinate (temp) at (#2);
	\ifthenelse{\equal{#1}{}}{
		\draw[line width=1pt,fill=#3] (temp)circle(.25);}
	{\ifthenelse{\equal{#1}{C}}{
			\draw[rounded corners=2pt,line width=1pt,fill=#3] ($(temp)-(.25,.25)$)rectangle++(.5,.5);}
		{\ifthenelse{\isodd{#1}}{
				\draw[rounded corners=2pt,line width=1pt,fill=#3] (temp)--++($(.25,0)+#1*(0,.25)$)--++($#1*(0,-.5)$)--++(-.5,0)--(temp);
            \coordinate (temp) at ($(temp)-#1*(0,.05)-(.05,0)$);}
			{\draw[rounded corners=2pt,line width=1pt,fill=#3] (temp)--++($(-.25,0)+#1*(0,.125)$)--++($#1*(0,-.25)$)--++(.5,0)--(temp);
            \coordinate (temp) at ($(temp)-#1*(0,.05)-(.05,0)$);}}}
	\node at (temp) {#4}}

\newcommand{\Gate}[5]{ 
	\coordinate (temp) at (#2);
	\ifthenelse{\equal{#1}{V}}{
		\draw[rounded corners=2pt,line width=1pt,fill=#4!50!white] ($(temp)-(.25,.25)$)rectangle++($(-.25,.5)+#3*(.75,0)$)node[midway]{#5}}
	{\draw[rounded corners=2pt,line width=1pt,fill=#4!50!white] ($(temp)-(.25,-.25)$)rectangle++($(.5,.25)-#3*(0,.75)$)node[midway]{#5}}}

\newcommand{\Qin}[2]{ 
	\coordinate (temp) at (#2);
	\StrLen {#1}[\tmpLn]
	\ifthenelse{ \tmpLn= 1}{
	\ifthenelse{\equal{#1}{V}}{
	\draw[line width=1pt, fill=yellow!75!red] ($(temp)-(0,.1)$)circle(.15)}
	{\draw[line width=1pt, fill=yellow!75!red] ($(temp)-(.1,0)$)circle(.15)}}
	{\ifthenelse{\equal{#1}{V0}}{
		\node[below] at (temp) {$|0\rangle$}}
		{\node[left,align=right] at (temp){$|0\rangle$}}}}

\newcommand{\Qout}[3]{ 
	\coordinate (temp) at (#2);
	\StrLen {#1}[\tmpLn]
	\ifthenelse{ \tmpLn= 1}{
	\ifthenelse{\equal{#1}{V}}{
		\draw[line width=1pt] ($(temp)+(-.1,.05)$) --++(.2,.1)node[above]{#3};
		\draw[line width=1pt] ($(temp)+(-.1,0)$) --++(.2,.1)}
	{\draw[line width=1pt] ($(temp)+(-.05,.1)$) --++(.1,-.2)node[right]{#3};
	\draw[line width=1pt] ($(temp)+(0,.1)$) --++(.1,-.2)}}{
	\ifthenelse{\equal{#1}{VM}}{\coordinate (temp2) at ($(temp)+(0,.25)$);}{\coordinate (temp2) at ($(temp)+(.25,0)$);}
	\draw[line width=1pt] ($(temp2)-(.25,.25)$)rectangle++(.5,.5);\draw[line width=1pt,->] ($(temp2)+(-.25,-.25)$)--++(.35,.35);}}

\newcommand{\Qreset}[3]{
	\coordinate (temp) at (#2);
	\ifthenelse{\equal{#1}{V}}{
		\ifthenelse{\equal{#3}{in}}{
			\draw[fill=white,color=white] ($(temp)-(.1,-.1)$)rectangle++(.2,.4);
			\Qout{V}{$(temp)$}{};
			\Qin{V}{$(temp)+(0,.5)$}}{
			\ifthenelse{\equal{#3}{out}}{\draw[fill=white,color=white] ($(temp)-(.1,-.1)$)rectangle++(.2,1.2);
				\Qout{VM}{$(temp)$}{};
				\Qin{V0}{$(temp)+(0,1.2)$}}{\draw[fill=white,color=white] ($(temp)-(.1,-.1)$)rectangle++(.2,.4);
				\Qout{V}{$(temp)$}{};
				\Qin{V0}{$(temp)+(0,.5)$}}}}{
		\ifthenelse{\equal{#3}{in}}{
			\draw[fill=white,color=white] ($(temp)-(0,.1)$)rectangle++(.4,.2);
			\Qout{H}{$(temp)$}{};
			\Qin{H}{$(temp)+(.5,0)$}}{
			\ifthenelse{\equal{#3}{out}}{
			\draw[fill=white,color=white] ($(temp)-(0,.1)$)rectangle++(1.1,.2);
			\Qout{HM}{$(temp)$}{};
			\Qin{H0}{$(temp)+(1.2,0)$}}{\draw[fill=white,color=white] ($(temp)-(0,.1)$)rectangle++(.65,.2);
			\Qout{H}{$(temp)$}{};
			\Qin{H0}{$(temp)+(.75,0)$}}}}}
	        

%% file: structure.tex
	\begin{tikzpicture}
		\coordinate (L1) at (0,0);
		\node at ($(L1)+(0,1)$) {(a)};
		
		\draw[line width=1pt] (L1)--++(0,-1.35);
		\draw[line width=1pt] ($(L1)+(.75,0)$)--++(0,-1.35);
		\draw[line width=1pt] ($(L1)+(1.5,0)$)--++(0,-1.35);
		\draw[line width=1pt] ($(L1)+(2.25,0)$)--++(0,-1.35);
		
		\draw[line width=1pt,fill=lblue,rounded corners=.5cm] ($(L1)+(-.25,.25)$)rectangle++(2.75,-1)node[midway] {$A^{\alpha\beta\gamma\delta}$ / $\langle\Psi|$};
		
		\node at ($(L1)+(3.5,-.5)$) {$\rightarrow$};
		
		\coordinate (L2) at ($(L1)+(5,-.5)$);
		
		\draw[line width=1pt] (L2)--($(L2)+(2,1)$)--($(L2)+(3,0)$)--($(L2)+(1,-.5)$)--($(L2)+(.5,.75)$)--($(L2)+(2,-.25)$)--(L2);
		\draw[line width=1pt] ($(L2)+(3,0)$)--(L2)--($(L2)+(1,-.5)$)--($(L2)+(2,1)$)--($(L2)+(.5,.75)$);
		\draw[line width=1pt] (L2)--($(L2)+(0,-.85)$);	
		\draw[line width=1pt] ($(L2)+(3,0)$)--($(L2)+(3,-.85)$);	
		\draw[line width=1pt] ($(L2)+(1,-.5)$)--($(L2)+(1,-.85)$);	
		\draw[line width=1pt] ($(L2)+(2,-.25)$)--($(L2)+(2,-.85)$);	
		
		\TNode{}{L2}{lblue}{};
		\TNode{}{$(L2)+(2,1)$}{lblue}{};
		\TNode{}{$(L2)+(3,0)$}{lblue}{};
		\TNode{}{$(L2)+(1,-.5)$}{lblue}{};
		\TNode{}{$(L2)+(.5,.75)$}{lblue}{};
		\TNode{}{$(L2)+(2,-.25)$}{lblue}{};
		
		\coordinate (MPS) at ($(L1)+(10.5,0)$);
		\node at ($(MPS)+(0,1)$) {(b)};
		
		\draw[line width=1pt] ($(MPS)-(0,.5)$)--(MPS)--++(3,0)--++(0,-.5);
		\draw[line width=1pt] ($(MPS)+(.75,0)$)--++(0,-.5);
		\draw[line width=1pt] ($(MPS)+(1.5,0)$)--++(0,-.5);
		\draw[line width=1pt] ($(MPS)+(2.25,0)$)--++(0,-.5);
		
		\TNode{}{MPS}{lblue}{};
		\TNode{}{$(MPS)+(.75,0)$}{lblue}{};
		\TNode{}{$(MPS)+(1.5,0)$}{lblue}{};
		\TNode{}{$(MPS)+(2.25,0)$}{lblue}{};
		\TNode{}{$(MPS)+(3,0)$}{lblue}{};

		\coordinate (PEPS) at ($(L1)-(0,3)$);
		\node at ($(PEPS)+(0,1)$) {(c)};
		
		\draw[line width=1pt] ($(PEPS)-(0,.5)$)--(PEPS)--++(1.5,0)--++(1,-2)--++(-1.5,0)--(PEPS);
		\draw[line width=1pt] ($(PEPS)+(.75,-.5)$)--++(0,.5)--++(1,-2)--++(0,-.5);
		\draw[line width=1pt] ($(PEPS)+(.5,-1.5)$)--++(0,.5)--++(1.5,0)--++(0,-.5);
		\draw[line width=1pt] ($(PEPS)+(1.5,0)$)--++(0,-.5);
		\draw[line width=1pt] ($(PEPS)+(1.25,-1)$)--++(0,-.5);
		\draw[line width=1pt] ($(PEPS)+(1,-2)$)--++(0,-.5);
		\draw[line width=1pt] ($(PEPS)+(2.5,-2)$)--++(0,-.5);
		
		\TNode{}{PEPS}{lblue}{};
		\TNode{}{$(PEPS)+(.75,0)$}{lblue}{};
		\TNode{}{$(PEPS)+(1.5,0)$}{lblue}{};
		\TNode{}{$(PEPS)+(.5,-1)$}{lblue}{};
		\TNode{}{$(PEPS)+(1.25,-1)$}{lblue}{};
		\TNode{}{$(PEPS)+(2,-1)$}{lblue}{};
		\TNode{}{$(PEPS)+(1,-2)$}{lblue}{};
		\TNode{}{$(PEPS)+(1.75,-2)$}{lblue}{};
		\TNode{}{$(PEPS)+(2.5,-2)$}{lblue}{};
		
		\coordinate (TTN) at ($(PEPS)+(4,0)$);
		\node at ($(TTN)+(0,1)$) {(d)};
		
		\draw[line width=1pt] ($(TTN)+(-.25,-2)$)--++(.25,.5)--++(.5,.75)--++(1,.75)--++(1,-.75)--++(.5,-.75)--++(.25,-.5);
		\draw[line width=1pt] ($(TTN)+(.75,-2)$)--++(.25,.5)--++(-.5,.75);
		\draw[line width=1pt] ($(TTN)+(1.75,-2)$)--++(.25,.5)--++(.5,.75);
		\draw[line width=1pt] ($(TTN)+(0,-1.5)$)--++(.25,-.5);
		\draw[line width=1pt] ($(TTN)+(1,-1.5)$)--++(.25,-.5);
		\draw[line width=1pt] ($(TTN)+(2,-1.5)$)--++(.25,-.5);
		\draw[line width=1pt] ($(TTN)+(3,-1.5)$)--++(-.25,-.5);
		
		\TNode{}{$(TTN)+(1.5,0)$}{lblue}{};
		\TNode{}{$(TTN)+(.5,-.75)$}{lblue}{};
		\TNode{}{$(TTN)+(2.5,-.75)$}{lblue}{};
		\TNode{}{$(TTN)+(0,-1.5)$}{lblue}{};
		\TNode{}{$(TTN)+(1,-1.5)$}{lblue}{};
		\TNode{}{$(TTN)+(2,-1.5)$}{lblue}{};
		\TNode{}{$(TTN)+(3,-1.5)$}{lblue}{};
		
		\coordinate (MERA) at ($(TTN)+(5,1)$);
		\node at ($(MERA)+(0,0)$) {(e)};
		
		\draw[line width=1pt] ($(MERA)+(0,-1.5)$)--($(MERA)+(.75,-.75)$)--($(MERA)+(2.25,0)$);
		\draw[line width=1pt] ($(MERA)+(1.5,-1.5)$)--($(MERA)+(.75,-.75)$)--++(-.5,.5);
		\draw[line width=1pt] ($(MERA)+(3,-1.5)$)--($(MERA)+(3.75,-.75)$)--($(MERA)+(2.25,0)$);
		\draw[line width=1pt] ($(MERA)+(4.5,-1.5)$)--($(MERA)+(3.75,-.75)$)--($(MERA)+(5.25,0)$)--++(.5,-.5);
		
		\draw[line width=1pt] ($(MERA)+(-.5,-2)$)--($(MERA)+(0,-1.5)$); 
		\draw[line width=1pt] ($(MERA)+(0,-3.75)+(.375,.75)$)--($(MERA)+(.75,-2.25)$)--($(MERA)+(0,-1.5)$); \draw[line width=1pt]($(MERA)+(.75,-3.75)+(.375,.75)$)--($(MERA)+(.75,-2.25)$)--($(MERA)+(1.5,-1.5)$);
		\draw[line width=1pt] ($(MERA)+(1.5,-3.75)+(.375,.75)$)--($(MERA)+(2.25,-2.25)$)--($(MERA)+(1.5,-1.5)$);
		\draw[line width=1pt] ($(MERA)+(2.25,-3.75)+(.375,.75)$)--($(MERA)+(2.25,-2.25)$)--($(MERA)+(3,-1.5)$);
		\draw[line width=1pt] ($(MERA)+(3,-3.75)+(.375,.75)$)--($(MERA)+(3.75,-2.25)$)--($(MERA)+(3,-1.5)$);
		\draw[line width=1pt] ($(MERA)+(3.75,-3.75)+(.375,.75)$)--($(MERA)+(3.75,-2.25)$)--($(MERA)+(4.5,-1.5)$);
		\draw[line width=1pt] ($(MERA)+(4.5,-3.75)+(.375,.75)$)--($(MERA)+(5.25,-2.25)$)--($(MERA)+(4.5,-1.5)$);
		\draw[line width=1pt] ($(MERA)+(5.5,-2.75)$)--($(MERA)+(5.25,-2.25)$)--($(MERA)+(5.5,-1.75)$);

        \foreach \i in {0,...,7}{
        \draw[line width=1pt] ($(MERA)+(-.25,-4.25)+\i*(.75,0)$)--++(.75,1.25);
        \draw[line width=1pt] ($(MERA)+(.25,-4.25)+\i*(.75,0)$)--++(-.75,1.25);}
		
		\TNode{}{$(MERA)+(2.25,0)$}{lblue}{};
		\TNode{}{$(MERA)+(5.25,0)$}{lblue}{};
		
		\TNode{C}{$(MERA)+(.75,-.75)$}{lblue}{};
		\TNode{C}{$(MERA)+(3.75,-.75)$}{lblue}{};
		
		\TNode{}{$(MERA)+(0,-1.5)$}{lblue}{};
		\TNode{}{$(MERA)+(1.5,-1.5)$}{lblue}{};
		\TNode{}{$(MERA)+(3,-1.5)$}{lblue}{};
		\TNode{}{$(MERA)+(4.5,-1.5)$}{lblue}{};
		
	    \foreach \i in {0,...,3}{\TNode{C}{$(MERA)+(.75,-2.25)+\i*(1.5,0)$}{lblue}{};}
		
		\TNode{}{$(MERA)+(0,-3.75)+(.375,.75)$}{lblue}{};
		\TNode{}{$(MERA)+(.75,-3.75)+(.375,.75)$}{lblue}{};
		\TNode{}{$(MERA)+(1.5,-3.75)+(.375,.75)$}{lblue}{};
		\TNode{}{$(MERA)+(2.25,-3.75)+(.375,.75)$}{lblue}{};
		\TNode{}{$(MERA)+(3,-3.75)+(.375,.75)$}{lblue}{};
		\TNode{}{$(MERA)+(3.75,-3.75)+(.375,.75)$}{lblue}{};
		\TNode{}{$(MERA)+(4.5,-3.75)+(.375,.75)$}{lblue}{};
		
	    \foreach \i in {0,...,7}{\TNode{C}{$(MERA)+(0,-3.75)+\i*(.75,0)$}{lblue}{};}		
	\end{tikzpicture}

%% file: Class2QTN.tex
\begin{tikzpicture}[scale=.9]
	\draw[line width=2pt] (0,0)circle(5.5);
	
	\coordinate (L2) at ($(120:5)+(0,.5)$);
	\coordinate (L1) at ($(60:4)+(0,0)$);
	\coordinate (L3) at ($(0:4.5)+(0,0)$);
	\coordinate (Rd1) at (8,-2);
	\coordinate (U1) at ($(300:4)+(0,1.25)$);
	\coordinate (U2) at ($(240:4)+(-4.5,1.25)$);
	\coordinate (U3) at ($(180:4)+(-4,1.75)$);
	
	\draw[white,fill=white] ($(L2)-(1.2,.8)$)rectangle++(1.5,1.2);
	\draw[white,fill=white] ($(L1)-(.5,.8)$)rectangle++(3,2);
	\draw[white,fill=white] ($(L3)+(.5,-.4)$)rectangle++(1,1.6);
	\draw[white,fill=white] ($(U1)-(.5,3.2)$)rectangle++(4,3.4);
	\draw[white,fill=white] ($(U2)+(1,.3)$)rectangle++(4.8,-4);
	\draw[white,fill=white] ($(U3)+(2,.3)$)rectangle++(1,-2.25);
	
	\node at ($(L2)+(-1.25,.15)$) {(a)};
	\draw[line width=1pt] (L2)--++(0,.75)node[right]{$l$};
	\TNode{}{L2}{lblue}{};
	\node at ($(L2)+(-.5,-.5)$) {$f^l(x)$};

	\node[fill=white] at ($(L1)+(0,1.5)$) {(b)};
	
	\draw[line width=1pt] (L1)--++(0,.5);
	\draw[line width=1pt] ($(L1)+(.75,0)$)--++(0,.5);
	\draw[line width=1pt] ($(L1)+(1.5,0)$)--++(0,.5);
	\draw[line width=1pt] ($(L1)+(2.25,0)$)--++(0,.5);
	\draw[line width=1pt] ($(L1)+(3,0)$)--++(0,.5);
	\draw[line width=1pt] ($(L1)+(3.75,0)$)--++(0,1.5)node[right]{$l$};
	
	\draw[rounded corners=.25cm,fill=lblue,line width=1pt]  ($(L1)+(-.25,.5)$)rectangle++(4.25,.5);
	
	\TNode{}{L1}{orng}{};
	\TNode{}{$(L1)+(.75,0)$}{orng}{};
	\TNode{}{$(L1)+(1.5,0)$}{orng}{};
	\TNode{}{$(L1)+(2.25,0)$}{orng}{};
	\TNode{}{$(L1)+(3,0)$}{orng}{};
	\TNode{}{$(L1)+(3.75,0)$}{orng}{};
	
	\node at ($(L1)+(4.5,0)$) {$\Phi(x)$};
	\node at ($(L1)+(4.5,.75)$) {$W^l$};
	
	\node[fill=white] at ($(L3)+(0,1.25)$) {(c)};
	\node at ($(L3)+(2.5,1.25)$) {$\chi=4$};
	
	\draw[line width=1pt] (L3)--++(0,.5)--++(3.75,0);
	\draw[line width=1pt] ($(L3)+(.75,0)$)--++(0,.5);
	\draw[line width=1pt] ($(L3)+(1.5,0)$)--++(0,.5);
	\draw[line width=1pt] ($(L3)+(2.25,0)$)--++(0,.5);
	\draw[line width=1pt] ($(L3)+(3,0)$)--++(0,.5);
	\draw[line width=1pt] ($(L3)+(3.75,0)$)--++(0,1)node[right]{$l$};
	
	\foreach \i in {0,...,5}{\TNode{-1}{$(L3)+(0,.5)+\i*(.75,0)$}{lblue}{};
		\TNode{}{$(L3)+\i*(.75,0)$}{orng}{};}

	\node at ($(U1)+(0,.5)$) {(d)};	
	\foreach \i in {0,...,5}{
	\draw[thick] ($(U1)-\i*(0,.75)$)--++(4,0);
	\Qin{H}{$(U1)-\i*(0,.75)$};}
	\foreach \i in {0,...,4}{
	\Qout{H}{$(U1)+(4,0)-\i*(0,.75)$}{};}
	\Qout{HM}{$(U1)+(4,-3.75)$}{};
	\Gate{H}{$(U1)+(.5,0)$}{2}{lblue}{};
	\foreach \i in {0,...,3}{
	\Gate{H}{$(U1)+(1.25,0)+\i*(.75,-.75)$}{3}{lblue}{};}

	\node at ($(U2)+(0,.5)$) {(e)};	
	\foreach \i in {0,...,5}{
		\draw[thick] ($(U2)-\i*(0,.75)$)--++(5.5,0);
		\Qin{H}{$(U2)-\i*(0,.75)$};}
	\foreach \i in {0,...,4}{
		\Qout{H}{$(U2)+(5.5,0)-\i*(0,.75)$}{};
		\Gate{H}{$(U2)+(.5,0)+\i*(.75,-.75)$}{2}{lblue}{};
		\Gate{H}{$(U2)+(2,0)+\i*(.75,-.75)$}{2}{lblue}{};}
	\Qout{HM}{$(U2)+(5.5,-3.75)$}{};

	\node at ($(U3)+(0,.5)$) {(f)};
	
	\draw[thick] (U3)--++(8.5,0);
	\draw[thick] ($(U3)-(0,1.5)$)--++(8.5,0);
	\Gate{H}{$(U3)+(3.5,0)$}{3}{lblue}{};
	\draw[fill=white,white] ($(U3)+(3.2,-.5)$)rectangle++(0.6,-.5);
	\draw[thick] ($(U3)-(0,.75)$)--++(8.5,0);
	\foreach \i in {0,...,2}{\Qin{H}{$(U3)-\i*(0,.75)$};}
	
	\Gate{H}{$(U3)+(.5,0)$}{2}{lblue}{};
	\Gate{H}{$(U3)+(1.25,-.75)$}{2}{lblue}{};
	\Gate{H}{$(U3)+(2,0)$}{2}{lblue}{};
	\Qreset{H}{$(U3)+(2.5,0)$}{in};
	
	\Gate{H}{$(U3)+(4.25,-.75)$}{2}{lblue}{};
	
	\Gate{H}{$(U3)+(5,0)$}{3}{lblue}{};
	\draw[fill=white,white] ($(U3)+(4.7,-.5)$)rectangle++(0.6,-.5);
	\Qreset{H}{$(U3)+(4.75,-.75)$}{in};
	\Qreset{H}{$(U3)+(5.5,-1.5)$}{in};
	
	\Gate{H}{$(U3)+(5.75,0)$}{2}{lblue}{};
	\Gate{H}{$(U3)+(6.5,-.75)$}{2}{lblue}{};
	\Gate{H}{$(U3)+(7.25,0)$}{2}{lblue}{};
	\Gate{H}{$(U3)+(8,-.75)$}{2}{lblue}{};
	
	\Qout{H}{$(U3)+(8.5,0)$}{};
	\Qout{H}{$(U3)+(8.5,-.75)$}{};
	\Qout{HM}{$(U3)+(8.5,-1.5)$}{};

	\draw[thick,fill=white!80!black,rounded corners=5pt] ($(Rd1)-(.875,.875)$)rectangle++(1.75,1.75);
	\draw[thick] ($(Rd1)-(.5,0)$)node[above]{$\alpha$}--++(1,0)node[above]{$\beta$};
	\draw[thick] (Rd1)--++(0,-.5)node[right]{$i$};
	\TNode{-1}{Rd1}{lblue}{};
	
	\coordinate (URd1) at ($(Rd1)-(.375,2.5)$);
	\draw[thick,fill=white!80!black,rounded corners=5pt] ($(URd1)-(.5,.75)$)rectangle++(1.75,1.75);
	\Qout{V}{$(URd1)+(0,.5)$}{};
	\draw[thick] ($(URd1)+(0,.5)$)--++(0,-1)node[right]{$\alpha$};
	\draw[thick] ($(URd1)+(.75,.5)$)node[right]{$\beta$}--++(0,-1)node[right]{$i$};
	\Gate{V}{URd1}{2}{lblue}{${U^\dag}$};
\end{tikzpicture}

%% file: TN2QTN.tex
\begin{tikzpicture}[scale=1.2]
		\node at (0,1) {(a)};
		
		
		
		
		\node at (0,0) {$R^{\beta(k)}_{i\alpha}$};
		\coordinate (R1) at (1.5,0);
		\draw[thick] ($(R1)-(.5,0)$)node[above]{$\alpha$}--++(1,0)node[above]{$\beta$};
		\draw[thick] (R1)--++(0,.5)node[right]{$i$};
		\draw[rounded corners=2pt,line width=1pt,fill=lblue] (R1)--++(.25,.25)--++(0,-.5)--++(-.5,0)--(R1);
		
		\coordinate (UR1) at (3,0);
		\draw[thick] ($(UR1)-(0,.5)$)node[below]{$|0\rangle $}--++(0,1)node[above]{$\alpha$};
		\draw[thick] ($(UR1)-(-.5,.5)$)node[below]{$\beta$}--++(0,1)node[above]{$i$};
		\draw[rounded corners=2pt,line width=1pt,fill=lblue!50!white] ($(UR1)-(.25,.25)$)rectangle++(1,.5)node[midway]{$U^{(k)}_R$};
		
		\node at (5,0) {$\langle\alpha i| U^{(k)}_R | 0\beta\rangle$};

		\node at (0,-2) {${R^\dag}^{i\alpha(k)}_{\beta}$};
		\coordinate (Rd1) at (1.5,-2);
		\draw[thick] ($(Rd1)-(.5,0)$)node[above]{$\alpha$}--++(1,0)node[above]{$\beta$};
		\draw[thick] (Rd1)--++(0,-.5)node[right]{$i$};
		\draw[rounded corners=2pt,line width=1pt,fill=lblue] (Rd1)--++(.25,-.25)--++(0,.5)--++(-.5,0)--(Rd1);
		
		\coordinate (URd1) at (3,-2);
		\draw[thick] ($(URd1)+(0,.5)$)--++(0,-1)node[below]{$\alpha$};
		\draw[thick] ($(URd1)+(-.1,.45)$) --++(.2,.1);
		\draw[thick] ($(URd1)+(-.1,.5)$)node[above]{*} --++(.2,.1);
		\draw[thick] ($(URd1)+(.5,.5)$)node[above]{$\beta$}--++(0,-1)node[below]{$i$};
		\draw[rounded corners=2pt,line width=1pt,fill=lblue!50!white] ($(URd1)-(.25,.25)$)rectangle++(1,.5)node[midway]{${U^\dag}^{(k)}_R$};
		
		\node at (5,-2) {$\langle\beta 0| {U^\dag}^{(k)}_R | i\alpha \rangle$};
		
		\node at (7,1) {(b)};
		
		\draw[thick] (7.5,0)--++(-.5,0)--++(0,-1)--++(.5,0);
		\node at (7.75,-.5) {$=$};
		
		\coordinate (Rd2) at (8.5,0);
		\coordinate (R2) at (8.5,-1);
		\draw[thick] ($(R2)+(.5,0)$)node[above]{$\beta$}--++(-1,0)--++(0,1)--++(1,0)node[above]{$\beta'$};
		\draw[thick] (R2)--(Rd2);
		\draw[rounded corners=2pt,line width=1pt,fill=lblue] (R2)--++(.25,.25)--++(0,-.5)--++(-.5,0)--(R2);
		
		\coordinate (UR2) at (10,-1);
		\coordinate (URd2) at (10,0);
		\draw[thick] ($(UR2)-(0,.5)$)node[below]{$|0\rangle $}--++(0,1);
		\draw[thick] ($(UR2)-(-.5,.5)$)node[below]{$\beta$}--++(0,1);
		\draw[rounded corners=2pt,line width=1pt,fill=lblue!50!white] ($(UR2)-(.25,.25)$)rectangle++(1,.5)node[midway]{$U^{(k)}_R$};

		\draw[rounded corners=2pt,line width=1pt,fill=lblue] (Rd2)--++(.25,-.25)--++(0,.5)--++(-.5,0)--(Rd2);
		
		\draw[thick] ($(URd2)+(0,.5)$)--++(0,-1);
		\draw[thick] ($(URd2)+(-.1,.45)$) --++(.2,.1);
		\draw[thick] ($(URd2)+(-.1,.5)$)node[above]{*} --++(.2,.1);
		\draw[thick] ($(URd2)+(.5,.5)$)node[above]{$\beta'$}--++(0,-1);
		\draw[rounded corners=2pt,line width=1pt,fill=lblue!50!white] ($(URd2)-(.25,.25)$)rectangle++(1,.5)node[midway]{${U^\dag}^{(k)}_R$};
		
		\node at (11,-.5){$=$};
		\draw[thick] (11.5,.5)--++(0,-2)node[below]{$|0\rangle $};
		\draw[thick] (12,.5)node[above]{$\beta'$}--++(0,-2)node[below]{$\beta$};
		\draw[thick] (11.4,.45) --++(.2,.1);
		\draw[thick] (11.4,.5)node[above]{*} --++(.2,.1);
	\end{tikzpicture}

%% file: MPSlayouts.tex
\begin{tikzpicture}
	\node at (-1,1) {(a)};
	
	\coordinate (C1) at (0,0);
 
	\coordinate (U4) at ($(C1)-(-1,6)$);
	
	\draw[line width=1pt] ($(C1)-(.5,0)$)--(C1)--++(0,-4.5);
	\draw[line width=1pt] ($(C1)-(.5,.75)$)--++(.5,0);
	\draw[line width=1pt] ($(C1)-(.5,1.5)$)--++(.5,0);
	\draw[line width=1pt] ($(C1)-(0,2.25)$)--++(.5,0)node[above]{$l$};
	\draw[line width=1pt] ($(C1)-(.5,3)$)--++(.5,0);
	\draw[line width=1pt] ($(C1)-(.5,3.75)$)--++(.5,0);
	\draw[line width=1pt] ($(C1)-(.5,4.5)$)--++(.5,0);
	
	\TNode{}{$(C1)-(0,2.25)$}{red}{};

    \foreach \i in{0,...,2}{\TNode{}{$(C1)-(0.5,0)-\i*(0,.75)$}{orng}{};
    \TNode{}{$(C1)-(0.5,3)-\i*(0,.75)$}{orng}{};
    \TNode{1}{$(C1)-\i*(0,.75)$}{lblue}{};
    \TNode{-1}{$(C1)-(0,3)-\i*(0,.75)$}{lblue}{};}

	\coordinate (U3) at ($(C1)+(1.5,0)$);
	\draw[line width= 1pt] (U3)--++(2.5,0);
	\Qin{H}{U3};
	\draw[line width= 1pt] ($(U3)-(0,0.75)$)--++(2.5,0);
	\Qin{H}{$(U3)-(0,0.75)$};
	\draw[line width= 1pt] ($(U3)-(0,1.5)$)--++(2.5,0);
	\Qin{H}{$(U3)-(0,1.5)$};
	\draw[line width= 1pt] ($(U3)-(0,2.25)$)--++(2.5,0);
	\Qin{H}{$(U3)-(0,2.25)$};
	\draw[line width= 1pt] ($(U3)-(0,3)$)--++(2.5,0);
	\Qin{H}{$(U3)-(0,3)$};
	\draw[line width= 1pt] ($(U3)-(0,3.75)$)--++(2.5,0);
	\Qin{H}{$(U3)-(0,3.75)$};
	
	\Gate{H}{$(U3)+(.5,0)$}{2}{lblue}{};
	\Gate{H}{$(U3)+(.5,-3)$}{2}{lblue}{};
	\Gate{H}{$(U3)+(1.25,-.75)$}{2}{lblue}{};
	\Gate{H}{$(U3)+(1.25,-2.25)$}{2}{lblue}{};
	\Gate{H}{$(U3)+(2,-1.5)$}{2}{red}{};
	
	\Qout{H}{$(U3)+(2.5,0)$}{};
	\Qout{H}{$(U3)+(2.5,-.75)$}{};
	\Qout{HM}{$(U3)+(2.5,-1.5)$}{};
	\Qout{H}{$(U3)+(2.5,-2.25)$}{};
	\Qout{H}{$(U3)+(2.5,-3)$}{};
	\Qout{H}{$(U3)+(2.5,-3.75)$}{};
	
	\draw[line width= 1pt] (U4)--++(2.5,0);
	\draw[line width= 1pt] ($(U4)-(0,0.75)$)--++(3.25,0);
	\draw[line width= 1pt] ($(U4)-(0,1.5)$)--++(3.25,0);
	\draw[line width= 1pt] ($(U4)-(0,2.25)$)--++(2.5,0);
	
	\Qin{H}{U4};
	\Qin{H}{$(U4)-(0,.75)$};
	\Qreset{H}{$(U4)+(1,0)$}{in};
	\Qin{H}{$(U4)-(0,1.5)$};
	\Qin{H}{$(U4)-(0,2.25)$};
	\Qreset{H}{$(U4)+(1,-2.25)$}{in};
	
	\Gate{H}{$(U4)+(.5,0)$}{2}{lblue}{};
	\Gate{H}{$(U4)+(2,0)$}{2}{lblue}{};
	\Gate{H}{$(U4)+(.5,-1.5)$}{2}{lblue}{};
	\Gate{H}{$(U4)+(2,-1.5)$}{2}{lblue}{};
	\Gate{H}{$(U4)+(2.75,-.75)$}{2}{red}{};
	
	\Qout{H}{$(U4)+(2.5,0)$}{};
	\Qout{HM}{$(U4)+(3.25,-.75)$}{};
	\Qout{H}{$(U4)+(3.25,-1.5)$}{};
	\Qout{H}{$(U4)+(2.5,-2.25)$}{};
	
	\coordinate (L1) at ($(C1)+(7,0)$);
    \node at ($(L1)+(-1,1)$) {(b)};
	\draw[thick] (L1)--++(0,-.75);
	\draw[thick] ($(L1)-(.5,0)$)--++(.5,0);
	\draw[thick] ($(L1)-(.5,.75)$)--++(1.5,0)--++(0,-1.5);
	\draw[rounded corners=2pt,line width=1pt,fill=lblue] ($(L1)-(0,.75)$)--++(.25,.25)--++(0,-.5)--++(-.5,0)--($(L1)-(0,.75)$);
	\draw[thick] ($(L1)+(0,-1.5)$)--++(0,-.75);
	\draw[thick] ($(L1)-(.5,1.5)$)--++(.5,0);
	\draw[thick] ($(L1)-(.5,2.25)$)--++(1.5,0)--++(0,-1.5);
	\draw[rounded corners=2pt,line width=1pt,fill=lblue] ($(L1)-(0,2.25)$)--++(.25,.25)--++(0,-.5)--++(-.5,0)--($(L1)-(0,2.25)$);
	\draw[rounded corners=2pt,line width=1pt,fill=lblue] ($(L1)-(-1,2.25)$)--++(.25,.25)--++(0,-.5)--++(-.5,0)--($(L1)-(-1,2.25)$);
	\draw[thick] ($(L1)+(0,-3)$)--++(0,-.75);
	\draw[thick] ($(L1)-(.5,3)$)--++(.5,0);
	\draw[thick] ($(L1)-(.5,3.75)$)--++(2,0)node[above]{$l$};
	\draw[rounded corners=2pt,line width=1pt,fill=lblue] ($(L1)-(0,3.75)$)--++(.25,.25)--++(0,-.5)--++(-.5,0)--($(L1)-(0,3.75)$);
	\draw[rounded corners=2pt,line width=1pt,fill=lblue] ($(L1)-(-1,3.75)$)--++(.25,.25)--++(0,-.5)--++(-.5,0)--($(L1)-(-1,3.75)$);
    \foreach \i in{0,...,5}{\TNode{}{$(L1)-\i*(0,.75)-(.5,0)$}{orng}{};}

	\coordinate (U1) at ($(L1)+(2.5,0)$);
	\draw[thick] (U1)--++(1,0);
	\draw[line width=1pt, fill=yellow!75!red] ($(U1)-(.1,0)$)circle(.15);
	\draw[thick] ($(U1)+(.95,.1)$)--++(.1,-.2);\draw[thick] ($(U1)+(1,.1)$)--++(.1,-.2);
	\draw[thick] ($(U1)-(0,.75)$)--++(2,0);
	\draw[line width=1pt, fill=yellow!75!red] ($(U1)-(.1,.75)$)circle(.15);
	\draw[rounded corners=2pt,line width=1pt,fill=lblue!50!white] ($(U1)+(.25,.25)$)rectangle++(.5,-1.25);
	\draw[thick] ($(U1)+(1.95,-.65)$)--++(.1,-.2);\draw[thick] ($(U1)+(2,-.65)$)--++(.1,-.2);
	\draw[thick] ($(U1)-(0,1.5)$)--++(1,0);
	\draw[line width=1pt, fill=yellow!75!red] ($(U1)-(.1,1.5)$)circle(.15);
	\draw[thick] ($(U1)+(.95,-1.4)$)--++(.1,-.2);\draw[thick] ($(U1)+(1,-1.4)$)--++(.1,-.2);
	\draw[thick] ($(U1)-(0,2.25)$)--++(3,0);
	\draw[line width=1pt, fill=yellow!75!red] ($(U1)-(.1,2.25)$)circle(.15);
	\draw[rounded corners=2pt,line width=1pt,fill=lblue!50!white] ($(U1)+(.25,.25)+(0,-1.5)$)rectangle++(.5,-1.25);
	\draw[rounded corners=2pt,line width=1pt,fill=lblue!50!white] ($(U1)+(.25,.25)+(1,-.75)$)rectangle++(.5,-2);
	\draw[thick] ($(U1)+(2.95,-2.15)$)--++(.1,-.2);\draw[thick] ($(U1)+(3,-2.15)$)--++(.1,-.2);
	\draw[thick] ($(U1)-(0,3)$)--++(1,0);
	\draw[line width=1pt, fill=yellow!75!red] ($(U1)-(.1,3)$)circle(.15);
	\draw[thick] ($(U1)+(.95,-2.9)$)--++(.1,-.2);\draw[thick] ($(U1)+(1,-2.9)$)--++(.1,-.2);
	\draw[thick] ($(U1)-(0,3.75)$)--++(3,0);
	\draw[line width=1pt, fill=yellow!75!red] ($(U1)-(.1,3.75)$)circle(.15);
	\draw[rounded corners=2pt,line width=1pt,fill=lblue!50!white] ($(U1)+(.25,.25)+(0,-3)$)rectangle++(.5,-1.25);
	\draw[rounded corners=2pt,line width=1pt,fill=lblue!50!white] ($(U1)+(.25,.25)+(1.75,-2.25)$)rectangle++(.5,-2);
	\draw[line width=1pt] ($(U1)+(3.5,-3.5)$)rectangle++(-.5,-.5);\draw[line width=1pt,->] ($(U1)+(3,-4)$)--++(.35,.35);
	
	\coordinate (U2) at ($(L1)+(0,-6)$);
	\draw[thick] (U2)--++(5.5,0);
	\draw[line width=1pt, fill=yellow!75!red] ($(U2)-(.1,0)$)circle(.15);
	\draw[line width=1pt] ($(U2)+(6,.25)$)rectangle++(-.5,-.5);\draw[line width=1pt,->] ($(U2)+(5.5,-.25)$)--++(.35,.35);
	\draw[thick] ($(U2)-(0,.75)$)--++(1,0);
	\draw[line width=1pt, fill=yellow!75!red] ($(U2)-(.1,.75)$)circle(.15);
	\draw[rounded corners=2pt,line width=1pt,fill=lblue!50!white] ($(U2)+(.25,.25)$)rectangle++(.5,-1.25);
	\draw[thick] ($(U2)+(.95,-.65)$)--++(.1,-.2);\draw[thick] ($(U2)+(1,-.65)$)--++(.1,-.2);
	\draw[thick] ($(U2)-(-1.5,.75)$)--++(1.75,0);
	\draw[thick] ($(U2)+(2.45,-1.4)$)--++(.1,-.2);\draw[thick] ($(U2)+(2.5,-1.4)$)--++(.1,-.2);
	\draw[line width=1pt, fill=yellow!75!red] ($(U2)-(-1.4,.75)$)circle(.15);
	\draw[thick] ($(U2)-(-1.5,1.5)$)--++(1,0);
	\draw[line width=1pt, fill=yellow!75!red] ($(U2)-(-1.4,1.5)$)circle(.15);
	\draw[rounded corners=2pt,line width=1pt,fill=lblue!50!white] ($(U2)+(1.75,-.5)$)rectangle++(.5,-1.25);
	\draw[thick] ($(U2)+(3.2,-.65)$)--++(.1,-.2);\draw[thick] ($(U2)+(3.25,-.65)$)--++(.1,-.2);
	\draw[rounded corners=2pt,line width=1pt,fill=lblue!50!white] ($(U2)+(2.5,.25)$)rectangle++(.5,-1.25);
	\draw[thick] ($(U2)-(-3.75,.75)$)--++(1.75,0);
	\draw[thick] ($(U2)+(4.7,-1.4)$)--++(.1,-.2);\draw[thick] ($(U2)+(4.75,-1.4)$)--++(.1,-.2);
	\draw[line width=1pt, fill=yellow!75!red] ($(U2)-(-3.65,.75)$)circle(.15);
	\draw[thick] ($(U2)-(-3.75,1.5)$)--++(1,0);
	\draw[line width=1pt, fill=yellow!75!red] ($(U2)-(-3.65,1.5)$)circle(.15);
	\draw[rounded corners=2pt,line width=1pt,fill=lblue!50!white] ($(U2)+(4,-.5)$)rectangle++(.5,-1.25);
	\draw[thick] ($(U2)+(5.45,-.65)$)--++(.1,-.2);\draw[thick] ($(U2)+(5.5,-.65)$)--++(.1,-.2);
	\draw[rounded corners=2pt,line width=1pt,fill=lblue!50!white] ($(U2)+(4.75,.25)$)rectangle++(.5,-1.25);

\end{tikzpicture}

%% file: TNLayouts.tex
	    \begin{tikzpicture}
	\node at (0,1) {(a)};
	
	\coordinate (U5) at (0,0);
	
	\draw[line width=1pt] (U5)--++(4.75,0);
	\draw[line width=1pt] ($(U5)+(0,-.75)$)--++(4.75,0);
	\draw[line width=1pt] ($(U5)+(0,-1.5)$)--++(4.75,0);
	\draw[line width=1pt] ($(U5)+(0,-2.25)$)--++(4.75,0);
	\draw[line width=1pt] ($(U5)+(0,-3)$)--++(4.75,0);
	\draw[line width=1pt] ($(U5)+(0,-3.75)$)--++(4.75,0);
	
	\Qin{H}{U5};
	\Qin{H}{$(U5)+(0,-.75)$};
	\Qin{H}{$(U5)+(0,-1.5)$};
	\Qin{H}{$(U5)+(0,-2.25)$};
	\Qin{H}{$(U5)+(0,-3)$};
	\Qin{H}{$(U5)+(0,-3.75)$};
	
	\foreach \i in {0,...,2}{\foreach \j in {0,...,2}{
		\Gate{H}{$(U5)+(.5,0)-\j*(0,1.5)+\i*(1.5,0)$}{2}{lblue}{};}
		\foreach \j in {0,1}{
		\Gate{H}{$(U5)+(1.25,-.75)-\j*(0,1.5)+\i*(1.5,0)$}{2}{lblue}{};}}	
	
	\Qout{H}{$(U5)+(4.75,0)$}{};
	\Qout{H}{$(U5)+(4.75,-.75)$}{};
	\Qout{HM}{$(U5)+(4.75,-1.5)$}{};
	\Qout{H}{$(U5)+(4.75,-2.25)$}{};
	\Qout{H}{$(U5)+(4.75,-3)$}{};
	\Qout{H}{$(U5)+(4.75,-3.75)$}{};
	
	\coordinate (U3)at($(U5)+(7,0)$);
	\node at ($(U3)+(0,1)$){(b)};
	
	\foreach \i in {0,...,5}{
	\draw[line width=1pt] ($(U3)+\i*(0,-.75)$)--++(4,0);
	\Qin{H}{$(U3)+\i*(0,-.75)$};}

	\Qout{H}{$(U3)+(4,0)$}{};
	\Qout{H}{$(U3)+(4,-.75)$}{};
	\Qout{H}{$(U3)+(4,-1.5)$}{};
	\Qout{H}{$(U3)+(4,-2.25)$}{};
	\Qout{HM}{$(U3)+(4,-3)$}{};
	\Qout{H}{$(U3)+(4,-3.75)$}{};
	
	\Gate{H}{$(U3)+(.5,-.75)$}{2}{lblue}{};
	\Gate{H}{$(U3)+(.5,-2.25)$}{2}{lblue}{};
	\Gate{H}{$(U3)+(1.25,0)$}{2}{lblue}{};
	\Gate{H}{$(U3)+(1.25,-1.5)$}{2}{lblue}{};
	\Gate{H}{$(U3)+(1.25,-3)$}{2}{lblue}{};
	\Gate{H}{$(U3)+(2,-.75)$}{4}{lblue}{};	
	\Gate{H}{$(U3)+(3.5,-.75)$}{4}{lblue}{};
	\draw[line width=1pt]($(U3)+(1.75,-1.5)$)--++(2,0);
	\draw[line width=1pt]($(U3)+(1.75,-2.25)$)--++(2,0);	
	\Gate{H}{$(U3)+(2.75,-.75)$}{2}{lblue}{};
	\Gate{H}{$(U3)+(2.75,-2.25)$}{2}{lblue}{};
	
	\coordinate (L2) at ($(U5)-(-.5,6)$);
	\node at ($(L2)+(-.5,1)$){(c)};
	
	\draw[line width=1pt] (L2)rectangle++(3,-3);
	\draw[line width=1pt] ($(L2)-(0,1.5)$)--++(3,0);
	\draw[line width=1pt] ($(L2)+(1.5,0)$)--++(0,-3);
	\draw[line width=1pt] ($(L2)+(3,-3)$)--++(.5,0)node[above]{$l$};
	
	\foreach \i in {1,...,3}{
	\foreach \j in {0,...,2}{\draw[line width=1pt]($(L2)+(0,1.5)+\i*(0,-1.5)+\j*(1.5,0)$)--++(-.5,.5);}
	\TNode{C}{$(L2)+(0,1.5)+\i*(0,-1.5)$}{lblue}{A\i};
	\TNode{C}{$(L2)+(1.5,1.5)+\i*(0,-1.5)$}{lblue}{B\i};
	\TNode{C}{$(L2)+(3,1.5)+\i*(0,-1.5)$}{lblue}{C\i};}

	\coordinate (U4) at ($(L2)+(6,.5)$);
	
	\draw[line width=1pt] ($(U4)+ (1.5,0)$)--++(2.5,0);
	\draw[line width=1pt] ($(U4)-(-.75,.75)$)--++(3.25,0);
	\draw[line width=1pt] ($(U4)-(0,1.5)$)--++(5.5,0);
	\draw[line width=1pt] ($(U4)-(-1.5,2.25)$)--++(4,0);
	\draw[line width=1pt] ($(U4)-(0,3)$)--++(5.5,0);
	\draw[line width=1pt] ($(U4)-(-.75,3.75)$)--++(3.25,0);
	\draw[line width=1pt] ($(U4)-(-1.5,4.5)$)--++(1,0);
	
	\Gate{H}{$(U4)+(.5,-1.5)$}{3}{lblue}{A1};
	\Qin{H0}{$(U4)+(0,-1.5)$};
	\Qin{H}{$(U4)+(0,-3)$};
	
	\Gate{H}{$(U4)+(1.25,-.75)$}{2}{lblue}{B1};
	\Gate{H}{$(U4)+(1.25,-3)$}{2}{lblue}{A2};
	\Qin{H}{$(U4)+(.75,-.75)$};
	\Qin{H}{$(U4)+(.75,-3.75)$};
	
	\Gate{H}{$(U4)+(2,0)$}{2}{lblue}{C1};
	\Gate{H}{$(U4)+(2,-1.5)$}{3}{lblue}{B2};
	\Gate{H}{$(U4)+(2,-3.75)$}{2}{lblue}{A3};
	\Qin{H}{$(U4)+(1.5,0)$};
	\Qin{H}{$(U4)+(1.5,-2.25)$};
	\Qin{H}{$(U4)+(1.5,-4.5)$};
	
	\Qreset{H}{$(U4)+(2.5,0)$}{in};
	\Qreset{H}{$(U4)+(2.5,-2.25)$}{in};
	\Qout{H}{$(U4)+(2.5,-4.5)$}{};
	
	\Gate{H}{$(U4)+(3.5,0)$}{3}{lblue}{C2};
	\Gate{H}{$(U4)+(3.5,-2.25)$}{3}{lblue}{B3};
	
	\Qreset{H}{$(U4)+(4,-2.25)$}{in};
	\Qout{H}{$(U4)+(4,0)$}{};
	\Qout{H}{$(U4)+(4,-.75)$}{};
	\Qout{H}{$(U4)+(4,-3.75)$}{};
	
	\Gate{H}{$(U4)+(5,-1.5)$}{3}{lblue}{C3};
	
	\Qout{HM}{$(U4)+(5.5,-1.5)$}{};
	\Qout{H}{$(U4)+(5.5,-2.25)$}{};
	\Qout{H}{$(U4)+(5.5,-3)$}{};
	
	
	
	
		
	

\end{tikzpicture}

%% file: QTNWall.tex
\begin{tikzpicture}
	\coordinate (L1) at (0,0);
	\node at ($(L1)+(0,1)$) {(a)};
	
	\draw[line width=1pt] ($(L1)+(.5,0)$)--(L1)--++(0,-2.25)--++(.5,0);
	\draw[line width=1pt] ($(L1)+(.5,-.75)$)--++(-.5,0);
	\draw[line width=1pt] ($(L1)+(.5,-1.5)$)--++(-.5,0);
	
	\TNode{-2}{L1}{lblue}{};
	\TNode{-2}{$(L1)-(0,.75)$}{lblue}{};
	\TNode{-2}{$(L1)-(0,1.5)$}{lblue}{};
	\TNode{-2}{$(L1)-(0,2.25)$}{lblue}{};
	
	\node at ($(L1)+(1,-1)$) {$\rightarrow$};
	
	\coordinate (U1) at ($(L1)+(2,0)$);
	
	\draw[line width=1pt] (U1)--++(2.75,0);
	\draw[line width=1pt] ($(U1)+(0,-.75)$)--++(2.75,0);
	\draw[line width=1pt] ($(U1)+(0,-1.5)$)--++(2.75,0);
	\draw[line width=1pt] ($(U1)+(0,-2.25)$)--++(2.75,0);
	
	\Gate{H}{$(U1)+(.5,0)$}{2}{white}{$I$};
	\Gate{H}{$(U1)+(.5,-1.5)$}{2}{lblue}{$U_c$};
	\Gate{H}{$(U1)+(1.25,-.75)$}{2}{lblue}{$U_c$};
	\Gate{H}{$(U1)+(2,0)$}{2}{lblue}{$U_c$};
	\Gate{H}{$(U1)+(2,-1.5)$}{2}{white}{$I$};
	
	\Qin{H0}{U1};
	\Qin{H0}{$(U1)-(0,.75)$};
	\Qin{H0}{$(U1)-(0,1.5)$};
	\Qin{H0}{$(U1)-(0,2.25)$};
	
	\Qout{HM}{$(U1)+(2.75,0)$};
	\Qout{HM}{$(U1)-(-2.75,.75)$};
	\Qout{HM}{$(U1)-(-2.75,1.5)$};
	\Qout{HM}{$(U1)-(-2.75,2.25)$};
	
	\node at ($(U1)+(3.75,-1)$) {$\rightarrow$};
	
	\coordinate (U2) at ($(U1)+(4.75,0)$);
	
	\draw[line width=1pt] (U2)--++(2.75,0);
	\draw[line width=1pt] ($(U2)+(0,-.75)$)--++(2.75,0);
	\draw[line width=1pt] ($(U2)+(0,-1.5)$)--++(2.75,0);
	\draw[line width=1pt] ($(U2)+(0,-2.25)$)--++(2.75,0);
	
	\Gate{H}{$(U2)+(.5,0)$}{2}{green!80!black}{$W$};
	\Gate{H}{$(U2)+(.5,-1.5)$}{2}{green!50!black}{$U_o$};
	\Gate{H}{$(U2)+(1.25,-.75)$}{2}{green!50!black}{$U_o$};
	\Gate{H}{$(U2)+(2,0)$}{2}{green!50!black}{$U_o$};
	\Gate{H}{$(U2)+(2,-1.5)$}{2}{green!80!black}{$W$};
	
	\Qin{H0}{U2};
	\Qin{H0}{$(U2)-(0,.75)$};
	\Qin{H0}{$(U2)-(0,1.5)$};
	\Qin{H0}{$(U2)-(0,2.25)$};
	
	\Qout{HM}{$(U2)+(2.75,0)$};
	\Qout{HM}{$(U2)-(-2.75,.75)$};
	\Qout{HM}{$(U2)-(-2.75,1.5)$};
	\Qout{HM}{$(U2)-(-2.75,2.25)$};
	
	\coordinate (L2) at ($(L1)-(-.5,4.5)$);
	\node at ($(L2)+(-.5,1)$) {(b)};
	
	\draw[double,line width=1pt] (L2)--++(7.75,0);
	
	\Qin{H0}{L2};
	\Gate{H}{$(L2)+(.5,0)$}{1}{dyell}{$U_R$};
	
	\draw[line width=2pt] ($(L2)+(1.25,.5)$)--++(-.25,0)--++(0,-1.75)--++(.25,0);
	\draw[line width=2pt] ($(L2)+(2.75,.5)$)--++(.25,0)node[right,align=left]{$N_b$}--++(0,-1.75)--++(-.25,0);
	\draw[line width=1pt] ($(L2)+(1.75,-.75)$)--++(1,0);
	\Qin{H0}{$(L2)+(1.75,-.75)$};
	\Gate{H}{$(L2)+(2.25,0)$}{2}{lblue}{$U_G$};
	\Qout{H}{$(L2)+(2.75,-.75)$}{};
	
	\draw[line width=2pt] ($(L2)+(4,.5)$)--++(-.25,0)--++(0,-1.75)--++(.25,0);
	\draw[line width=2pt] ($(L2)+(5.5,.5)$)--++(.25,0)node[right,align=left]{$N_x$}--++(0,-1.75)node[right,align=left]{$i=1$}--++(-.25,0);
	\draw[line width=1pt] ($(L2)+(4.5,-.75)$)--++(1,0);
	\node[left,align=right] at ($(L2)+(4.5,-.75)$){$|x_i\rangle$};
	\Gate{H}{$(L2)+(5,0)$}{2}{dgreen}{$U_D^{(i)}$};
	\Qout{H}{$(L2)+(5.5,-.75)$}{};
	
	\draw[line width=1pt] ($(L2)+(6.75,-.75)$)--++(1,0);
	\Qin{H0}{$(L2)+(6.75,-.75)$};
	\Gate{H}{$(L2)+(7.25,0)$}{2}{red}{$U_C$};
	\Qout{H}{$(L2)+(7.75,0)$}{};
	\Qout{HM}{$(L2)+(7.75,-.75)$}{};
	
\end{tikzpicture}

%% file: EncStrat.tex
\begin{tikzpicture}
    \coordinate (U1) at (0,0);
	\node at ($(U1)+(-.5,1)$) {(a)};

    \foreach \i in {0,...,5}{
	\draw[line width=1pt] ($(U1)-\i*(0,.75)$)--++(4,0);
    \Qin{H0}{$(U1)-\i*(0,.75)$};
    \Qout{HM}{$(U1)+(4,0)-\i*(0,.75)$}{};}
    \foreach \i in {0,...,4}{\Gate{H}{$(U1)+(.5,0)+\i*(.75,-.75)$}{2}{lblue}{$U^\dag_\i$};}

    \coordinate (U2) at ($(U1)+(6,0)$);

    \foreach \i in {0,...,5}{\Qout{HM}{$(U2)+(4,0)-\i*(0,.75)$}{};}
    
    \draw[line width=1pt] ($(U2)+(4,0)$)--++(-1,0);
    \draw[line width=1pt] ($(U2)+(4,-.75)$)--++(-2.5,0);
    \draw[line width=1pt] ($(U2)+(4,-1.5)$)--++(-1,0);
    \draw[line width=1pt] ($(U2)+(4,-2.25)$)--++(-3.25,0);
    \draw[line width=1pt] ($(U2)+(4,-3)$)--++(-1,0);
    \draw[line width=1pt] ($(U2)+(4,-3.75)$)--++(-3.25,0);
    
    \Gate{H}{$(U2)+(3.5,0)$}{2}{lblue}{$U^\dag_0$};
    \Gate{H}{$(U2)+(3.5,-1.5)$}{2}{lblue}{$U^\dag_1$};
    \Gate{H}{$(U2)+(3.5,-3)$}{2}{lblue}{$U^\dag_2$};
    \Gate{H}{$(U2)+(2,-.75)$}{3}{lblue}{$U^\dag_3$};
    \Gate{H}{$(U2)+(1.25,-2.25)$}{3}{lblue}{$U^\dag_4$};
    
    \Qin{H0}{$(U2)+(3,0)$};
    \Qin{H0}{$(U2)+(3,-1.5)$};
    \Qin{H0}{$(U2)+(3,-3)$};
    \Qin{H0}{$(U2)+(1.5,-.75)$};
    \Qin{H0}{$(U2)+(.75,-2.25)$};
    \Qin{H0}{$(U2)+(.75,-3.75)$};

    \coordinate (L1) at ($(U1)+(-.5,-6)$);
	\node at ($(L1)+ (0,1)$) {(b)};
	
	\draw[line width=1pt] (L1)rectangle++(.75,-2.25);
	\draw[line width=1pt] ($(L1)-(0,.75)$)rectangle++(.75,-.75);
	\draw[line width=1pt] ($(L1)-(0,3)$)rectangle++(.75,-2.25);
	\draw[line width=1pt] ($(L1)-(0,3.75)$)rectangle++(.75,-.75);
	\draw[line width=1pt] ($(L1)-(-.75,3)$)--++(0,.75);
	\draw[line width=1pt] ($(L1)-(-.75,5.25)$)--++(1.25,0)node[above]{$l$};
	
	\TNode{2}{L1}{lblue}{};
	\TNode{1}{$(L1)+(.75,0)$}{red}{};
	\TNode{2}{$(L1)+(0,-.75)$}{lblue}{};
	\TNode{1}{$(L1)+(.75,-.75)$}{red}{};
	\TNode{2}{$(L1)+(0,-1.5)$}{lblue}{};
	\TNode{1}{$(L1)+(.75,-1.5)$}{red}{};
	\TNode{2}{$(L1)+(0,-2.25)$}{lblue}{};
	\TNode{1}{$(L1)+(.75,-2.25)$}{red}{};
	\TNode{2}{$(L1)+(0,-3)$}{green!50!black}{};
	\TNode{1}{$(L1)+(.75,-3)$}{red}{};
	\TNode{2}{$(L1)+(0,-3.75)$}{green!50!black}{};
	\TNode{1}{$(L1)+(.75,-3.75)$}{red}{};
	\TNode{2}{$(L1)+(0,-4.5)$}{green!50!black}{};
	\TNode{1}{$(L1)+(.75,-4.5)$}{red}{};
	\TNode{2}{$(L1)+(0,-5.25)$}{green!50!black}{};
	\TNode{1}{$(L1)+(.75,-5.25)$}{red}{};
	\TNode{C}{$(L1)+(1.5,-5.25)$}{dyell}{};
	
	\coordinate (U5) at ($(L1)+(3.5,0)$);
	
	\draw[line width=1pt] (U5)--++(7,0);
	\draw[line width=1pt] ($(U5)+(0,-.75)$)--++(7,0);
	\draw[line width=1pt] ($(U5)+(0,-1.5)$)--++(7,0);
	\draw[line width=1pt] ($(U5)+(0,-2.25)$)--++(7,0);
	\draw[line width=1pt] ($(U5)+(0,-3)$)--++(7,0);
	\draw[line width=1pt] ($(U5)+(0,-3.75)$)--++(7,0);
	\draw[line width=1pt] ($(U5)+(0,-4.5)$)--++(7,0);
	\draw[line width=1pt] ($(U5)+(0,-5.25)$)--++(7.75,0);
	
	\Qin{H0}{U5};
	\Qin{H0}{$(U5)+(0,-.75)$};
	\Qin{H0}{$(U5)+(0,-1.5)$};
	\Qin{H0}{$(U5)+(0,-2.25)$};
	\Qin{H0}{$(U5)+(0,-3)$};
	\Qin{H0}{$(U5)+(0,-3.75)$};
	\Qin{H0}{$(U5)+(0,-4.5)$};
	\Qin{H0}{$(U5)+(0,-5.25)$};
	
	\Gate{H}{$(U5)+(.5,0)$}{2}{lblue}{};
	\Gate{H}{$(U5)+(1.25,-.75)$}{2}{lblue}{};
	\Gate{H}{$(U5)+(2,-1.5)$}{2}{lblue}{};
	\Gate{H}{$(U5)+(3.5,-3)$}{2}{green!50!black}{};
	\Gate{H}{$(U5)+(4.25,-3.75)$}{2}{green!50!black}{};
	\Gate{H}{$(U5)+(5,-4.5)$}{2}{green!50!black}{};
	
	\Gate{H}{$(U5)+(2,0)$}{2}{red}{};
	\Gate{H}{$(U5)+(2.75,-.75)$}{2}{red}{};
	\Gate{H}{$(U5)+(3.5,-1.5)$}{2}{red}{};
	\Gate{H}{$(U5)+(4.25,-2.25)$}{2}{red}{};
	\Gate{H}{$(U5)+(5,-3)$}{2}{red}{};
	\Gate{H}{$(U5)+(5.75,-3.75)$}{2}{red}{};
	\Gate{H}{$(U5)+(6.5,-4.5)$}{2}{red}{};
	\Gate{H}{$(U5)+(7.25,-5.25)$}{1}{yellow!75!black}{};	
	
	\Qout{H}{$(U5)+(7,0)$}{};
	\Qout{H}{$(U5)+(7,-.75)$}{};
	\Qout{H}{$(U5)+(7,-1.5)$}{};
	\Qout{H}{$(U5)+(7,-2.25)$}{};
	\Qout{H}{$(U5)+(7,-3)$}{};
	\Qout{H}{$(U5)+(7,-3.75)$}{};
	\Qout{H}{$(U5)+(7,-4.5)$}{};
	\Qout{HM}{$(U5)+(7.75,-5.25)$}{};
	
	\coordinate (I1) at ($(L1)-(0,7)$);
	\node at ($(I1)+(0,.5)$) {(c)};
 	
 	\foreach \i in {0,...,5}{
 	\foreach \j in {0,...,5}{
 		\pgfmathparse{70*rnd+30}
 		\edef\tmp{\pgfmathresult}
 		\draw[line width=1pt,fill=white!\tmp!black] ($(I1)-\i*(0,.5)+\j*(.5,0)$)rectangle++(.5,-.5);}}
 	\draw[line width=3pt,color=red] ($(I1)+(.25,-.25)$)--++(.5,0)--++(-.5,-.5)--++(0,-.5)--++(1,1)--++(.5,0)--++(-1.5,-1.5)--++(0,-.5)--++(2,2)--++(.5,0)--++(-2.5,-2.5)--++(.5,0)--++(2,2)--++(0,-.5)--++(-1.5,-1.5)--++(.5,0)--++(1,1)--++(0,-.5)--++(-.5,-.5)--++(.5,0);
 	
 	\coordinate (I2) at ($(I1)+(4,0)$);
 	
 	\foreach \i in {0,...,5}{
 		\foreach \j in {0,...,5}{
 			\pgfmathparse{70*rnd+30}
 			\edef\tmp{\pgfmathresult}
 			\draw[line width=1pt,fill=white!\tmp!black] ($(I2)-\i*(0,.5)+\j*(.5,0)$)rectangle++(.5,-.5);}}
 	\draw[line width=3pt,color=lblue] (I2)rectangle++(3,-1.5);
 	\draw[line width=3pt,color=lblue] ($(I2)+(0,-1.5)$)rectangle++(3,-1.5);
 	
 	\draw[line width=3pt,color=red] ($(I2)+(.25,-.25)$)--++(0,-1)--++(.5,0)--++(0,1)--++(.5,0)--++(0,-1)--++(.5,0)--++(0,1)--++(.5,0)--++(0,-1)--++(.5,0)--++(0,1);
 	\draw[line width=3pt,color=red] ($(I2)+(.25,-1.75)$)--++(0,-1)--++(.5,0)--++(0,1)--++(.5,0)--++(0,-1)--++(.5,0)--++(0,1)--++(.5,0)--++(0,-1)--++(.5,0)--++(0,1);
 	
 	\coordinate (I3) at ($(I2)+(4,0)$);
 	
 	\foreach \i in {0,...,5}{
 		\foreach \j in {0,...,5}{
 			\pgfmathparse{70*rnd+30}
 			\edef\tmp{\pgfmathresult}
 			\draw[line width=1pt,fill=white!\tmp!black] ($(I3)-\i*(0,.5)+\j*(.5,0)$)rectangle++(.5,-.5);}}
 	\foreach \i in {0,1}{
 		\foreach \j in {0,1}{
 	\draw[line width=3pt,color=lblue] ($(I3)-\i*(0,1.5)+\j*(1.5,0)$)rectangle++(1.5,-1.5);
 	\draw[line width=3pt,color=red] ($(I3)+(.25,-.25)-\i*(0,1.5)+\j*(1.5,0)$)--++(1,0)--++(0,-.5)--++(-1,0)--++(0,-.5)--++(1,0);
 }}
 	
\end{tikzpicture}

%% file: main.bbl
\begin{thebibliography}{100}
\urlstyle{rm}
\expandafter\ifx\csname url\endcsname\relax
  \def\url#1{\texttt{#1}}\fi
\expandafter\ifx\csname urlprefix\endcsname\relax\def\urlprefix{URL }\fi
\expandafter\ifx\csname doiprefix\endcsname\relax\def\doiprefix{DOI: }\fi
\providecommand{\bibinfo}[2]{#2}
\providecommand{\eprint}[2][]{\url{#2}}

\bibitem{Shor.1994}
\bibinfo{author}{Shor, P.~W.}
\newblock \bibinfo{title}{Algorithms for quantum computation: discrete
  logarithms and factoring}.
\newblock In \emph{\bibinfo{booktitle}{Proceedings 35th Annual Symposium on
  Foundations of Computer Science}}, \bibinfo{pages}{124--134},
  \doiprefix\doiurl{10.1109/SFCS.1994.365700} (\bibinfo{publisher}{{IEEE
  Comput. Soc. Press}}, \bibinfo{year}{1994}).

\bibitem{Caro.2022}
\bibinfo{author}{Caro, M.~C.} \emph{et~al.}
\newblock \bibinfo{journal}{\bibinfo{title}{Generalization in quantum machine
  learning from few training data}}.
\newblock {\emph{\JournalTitle{Nature communications}}}
  \textbf{\bibinfo{volume}{13}}, \bibinfo{pages}{4919},
  \doiprefix\doiurl{10.1038/s41467-022-32550-3} (\bibinfo{year}{2022}).

\bibitem{Abbas.2021}
\bibinfo{author}{Abbas, A.} \emph{et~al.}
\newblock \bibinfo{journal}{\bibinfo{title}{The power of quantum neural
  networks}}.
\newblock {\emph{\JournalTitle{Nature Computational Science}}}
  \textbf{\bibinfo{volume}{1}}, \bibinfo{pages}{403--409},
  \doiprefix\doiurl{10.1038/s43588-021-00084-1} (\bibinfo{year}{2021}).

\bibitem{Perrier.15.08.2021}
\bibinfo{author}{Perrier, E.}, \bibinfo{author}{Youssry, A.} \&
  \bibinfo{author}{Ferrie, C.}
\newblock \bibinfo{title}{Qdataset: Quantum datasets for machine learning}.

\bibitem{Boixo.2018}
\bibinfo{author}{Boixo, S.} \emph{et~al.}
\newblock \bibinfo{journal}{\bibinfo{title}{Characterizing quantum supremacy in
  near-term devices}}.
\newblock {\emph{\JournalTitle{Nature Physics}}} \textbf{\bibinfo{volume}{14}},
  \bibinfo{pages}{595--600}, \doiprefix\doiurl{10.1038/s41567-018-0124-x}
  (\bibinfo{year}{2018}).

\bibitem{White.1992}
\bibinfo{author}{White, S.~R.}
\newblock \bibinfo{journal}{\bibinfo{title}{Density matrix formulation for
  quantum renormalization groups}}.
\newblock {\emph{\JournalTitle{Physical Review Letters}}}
  \textbf{\bibinfo{volume}{69}} (\bibinfo{year}{1992}).

\bibitem{StellanOstlund.1995}
\bibinfo{author}{{Stellan Ostlund}} \& \bibinfo{author}{{Stefan Rommer}}.
\newblock \bibinfo{journal}{\bibinfo{title}{Thermodynamic limit of density
  matrix renormalization}}.
\newblock {\emph{\JournalTitle{Physical Review Letters}}}
  \textbf{\bibinfo{volume}{75}} (\bibinfo{year}{1995}).

\bibitem{Bridgeman.2017}
\bibinfo{author}{Bridgeman, J.~C.} \& \bibinfo{author}{Chubb, C.~T.}
\newblock \bibinfo{journal}{\bibinfo{title}{Hand-waving and interpretive dance:
  An introductory course on tensor networks}}.
\newblock {\emph{\JournalTitle{Journal of Physics A: Mathematical and
  Theoretical}}} \textbf{\bibinfo{volume}{50}}, \bibinfo{pages}{223001},
  \doiprefix\doiurl{10.1088/1751-8121/aa6dc3} (\bibinfo{year}{2017}).

\bibitem{Cirac.2021}
\bibinfo{author}{Cirac, I.}, \bibinfo{author}{Perez-Garcia, D.},
  \bibinfo{author}{Schuch, N.} \& \bibinfo{author}{Verstraete, F.}
\newblock \bibinfo{journal}{\bibinfo{title}{Matrix product states and projected
  entangled pair states: Concepts, symmetries, and theorems}}.
\newblock {\emph{\JournalTitle{Reviews of Modern Physics}}}
  \textbf{\bibinfo{volume}{93}}, \bibinfo{pages}{959},
  \doiprefix\doiurl{10.1103/RevModPhys.93.045003} (\bibinfo{year}{2021}).

\bibitem{Evenbly.2009}
\bibinfo{author}{Evenbly, G.} \& \bibinfo{author}{Vidal, G.}
\newblock \bibinfo{journal}{\bibinfo{title}{Algorithms for entanglement
  renormalization}}.
\newblock {\emph{\JournalTitle{Physical Review B}}}
  \textbf{\bibinfo{volume}{79}}, \doiprefix\doiurl{10.1103/PhysRevB.79.144108}
  (\bibinfo{year}{2009}).

\bibitem{Schollwock.2011}
\bibinfo{author}{Schollw{\"o}ck, U.}
\newblock \bibinfo{journal}{\bibinfo{title}{The density-matrix renormalization
  group in the age of matrix product states}}.
\newblock {\emph{\JournalTitle{Annals of Physics}}}
  \textbf{\bibinfo{volume}{326}}, \bibinfo{pages}{96--192},
  \doiprefix\doiurl{10.1016/j.aop.2010.09.012} (\bibinfo{year}{2011}).

\bibitem{Cichocki.2016}
\bibinfo{author}{Cichocki, A.} \emph{et~al.}
\newblock \bibinfo{journal}{\bibinfo{title}{Low-rank tensor networks for
  dimensionality reduction and large-scale optimization problems: Perspectives
  and challenges part 1}}.
\newblock {\emph{\JournalTitle{Foundations and Trends in Machine Learning}}}
  \textbf{\bibinfo{volume}{9}}, \bibinfo{pages}{249--429},
  \doiprefix\doiurl{10.1561/2200000059} (\bibinfo{year}{2016}).

\bibitem{Vanderstraeten.2019}
\bibinfo{author}{Vanderstraeten, L.}, \bibinfo{author}{Haegeman, J.} \&
  \bibinfo{author}{Verstraete, F.}
\newblock \bibinfo{journal}{\bibinfo{title}{Tangent-space methods for uniform
  matrix product states}}.
\newblock {\emph{\JournalTitle{SciPost Physics Lecture Notes}}}
  \doiprefix\doiurl{10.21468/SciPostPhysLectNotes.7} (\bibinfo{year}{2019}).

\bibitem{Orus.2019}
\bibinfo{author}{Or{\'u}s, R.}
\newblock \bibinfo{journal}{\bibinfo{title}{Tensor networks for complex quantum
  systems}}.
\newblock {\emph{\JournalTitle{Nature Reviews Physics}}}
  \textbf{\bibinfo{volume}{1}}, \bibinfo{pages}{538--550},
  \doiprefix\doiurl{10.1038/s42254-019-0086-7} (\bibinfo{year}{2019}).

\bibitem{Batselier.2022}
\bibinfo{author}{Batselier, K.}
\newblock \bibinfo{journal}{\bibinfo{title}{Low-rank tensor decompositions for
  nonlinear system identification: A tutorial with examples}}.
\newblock {\emph{\JournalTitle{IEEE Control Systems}}}
  \textbf{\bibinfo{volume}{42}}, \bibinfo{pages}{54--74},
  \doiprefix\doiurl{10.1109/MCS.2021.3122268} (\bibinfo{year}{2022}).

\bibitem{Cichocki.2016b}
\bibinfo{author}{Cichocki, A.} \emph{et~al.}
\newblock \bibinfo{journal}{\bibinfo{title}{Tensor networks for dimensionality
  reduction and large-scale optimizations. part 2 applications and future
  perspectives}}.
\newblock {\emph{\JournalTitle{Foundations and Trends in Machine Learning}}}
  \textbf{\bibinfo{volume}{9}}, \bibinfo{pages}{249--429},
  \doiprefix\doiurl{10.1561/2200000067} (\bibinfo{year}{2016}).

\bibitem{Levine.2023}
\bibinfo{author}{Levine, Y.}, \bibinfo{author}{Sharir, O.},
  \bibinfo{author}{Cohen, N.} \& \bibinfo{author}{Shashua, A.}
\newblock \bibinfo{title}{Bridging many-body quantum physics and deep learning
  via tensor networks}.
\newblock In \bibinfo{editor}{Grohs, P.} \& \bibinfo{editor}{Kutyniok, G.}
  (eds.) \emph{\bibinfo{booktitle}{Mathematical aspects of deep learning}},
  \bibinfo{pages}{439--474}, \doiprefix\doiurl{10.1017/9781009025096.012}
  (\bibinfo{publisher}{{Cambridge University Press}},
  \bibinfo{address}{Cambridge}, \bibinfo{year}{2023}).

\bibitem{Liu.15.05.2020}
\bibinfo{author}{Liu, D.}, \bibinfo{author}{Yao, Z.} \& \bibinfo{author}{Zhang,
  Q.}
\newblock \bibinfo{title}{Quantum-classical machine learning by hybrid tensor
  networks}.

\bibitem{Schuld.2021}
\bibinfo{author}{Schuld, M.}, \bibinfo{author}{Sweke, R.} \&
  \bibinfo{author}{Meyer, J.~J.}
\newblock \bibinfo{journal}{\bibinfo{title}{Effect of data encoding on the
  expressive power of variational quantum-machine-learning models}}.
\newblock {\emph{\JournalTitle{Physical Review A}}}
  \textbf{\bibinfo{volume}{103}},
  \doiprefix\doiurl{10.1103/PhysRevA.103.032430} (\bibinfo{year}{2021}).

\bibitem{TamaraG.KoldaandBrettW.BaderSandiaNationalLaboratories.2018}
\bibinfo{author}{{Tamara G. Kolda and Brett W. Bader, Sandia National
  Laboratories}}.
\newblock \bibinfo{title}{Tensor decompositions: Applications}.
\newblock In \bibinfo{editor}{Moitra, A.} (ed.)
  \emph{\bibinfo{booktitle}{Algorithmic Aspects of Machine Learning}},
  \bibinfo{pages}{48--70}, \doiprefix\doiurl{10.1017/9781316882177.005}
  (\bibinfo{publisher}{{Cambridge University Press}}, \bibinfo{year}{2018}).

\bibitem{Zaletel.2020}
\bibinfo{author}{Zaletel, M.~P.} \& \bibinfo{author}{Pollmann, F.}
\newblock \bibinfo{journal}{\bibinfo{title}{Isometric tensor network states in
  two dimensions}}.
\newblock {\emph{\JournalTitle{Physical review letters}}}
  \textbf{\bibinfo{volume}{124}}, \bibinfo{pages}{037201},
  \doiprefix\doiurl{10.1103/PhysRevLett.124.037201} (\bibinfo{year}{2020}).

\bibitem{Geng.2022}
\bibinfo{author}{Geng, C.}, \bibinfo{author}{Hu, H.-Y.} \&
  \bibinfo{author}{Zou, Y.}
\newblock \bibinfo{journal}{\bibinfo{title}{Differentiable programming of
  isometric tensor networks}}.
\newblock {\emph{\JournalTitle{Machine Learning: Science and Technology}}}
  \textbf{\bibinfo{volume}{3}}, \bibinfo{pages}{015020},
  \doiprefix\doiurl{10.1088/2632-2153/ac48a2} (\bibinfo{year}{2022}).

\bibitem{Zhou.2020}
\bibinfo{author}{Zhou, Y.}, \bibinfo{author}{Stoudenmire, E.~M.} \&
  \bibinfo{author}{Waintal, X.}
\newblock \bibinfo{journal}{\bibinfo{title}{What limits the simulation of
  quantum computers?}}
\newblock {\emph{\JournalTitle{Physical Review X}}}
  \textbf{\bibinfo{volume}{10}}, \doiprefix\doiurl{10.1103/PhysRevX.10.041038}
  (\bibinfo{year}{2020}).

\bibitem{Nguyen.21.04.2021}
\bibinfo{author}{Nguyen, T.} \emph{et~al.}
\newblock \bibinfo{title}{Tensor network quantum virtual machine for simulating
  quantum circuits at exascale}.

\bibitem{McCaskey.2018}
\bibinfo{author}{McCaskey, A.}, \bibinfo{author}{Dumitrescu, E.},
  \bibinfo{author}{Chen, M.}, \bibinfo{author}{Lyakh, D.} \&
  \bibinfo{author}{Humble, T.}
\newblock \bibinfo{journal}{\bibinfo{title}{Validating quantum-classical
  programming models with tensor network simulations}}.
\newblock {\emph{\JournalTitle{PloS one}}} \textbf{\bibinfo{volume}{13}},
  \bibinfo{pages}{e0206704}, \doiprefix\doiurl{10.1371/journal.pone.0206704}
  (\bibinfo{year}{2018}).

\bibitem{Guo.2020}
\bibinfo{author}{Guo, C.}, \bibinfo{author}{Modi, K.} \&
  \bibinfo{author}{Poletti, D.}
\newblock \bibinfo{journal}{\bibinfo{title}{Tensor network based machine
  learning of non-markovian quantum processes}}.
\newblock {\emph{\JournalTitle{Physical Review A}}}
  \textbf{\bibinfo{volume}{102}},
  \doiprefix\doiurl{10.1103/PhysRevA.102.062414} (\bibinfo{year}{2020}).

\bibitem{Pednault.16.10.2017}
\bibinfo{author}{Pednault, E.} \emph{et~al.}
\newblock \bibinfo{title}{Pareto-efficient quantum circuit simulation using
  tensor contraction deferral}.

\bibitem{Barratt.2021}
\bibinfo{author}{Barratt, F.} \emph{et~al.}
\newblock \bibinfo{journal}{\bibinfo{title}{Parallel quantum simulation of
  large systems on small nisq computers}}.
\newblock {\emph{\JournalTitle{npj Quantum Information}}}
  \textbf{\bibinfo{volume}{7}}, \doiprefix\doiurl{10.1038/s41534-021-00420-3}
  (\bibinfo{year}{2021}).

\bibitem{Jaschke.24.05.2022}
\bibinfo{author}{Jaschke, D.} \& \bibinfo{author}{Montangero, S.}
\newblock \bibinfo{title}{Is quantum computing green? an estimate for an
  energy-efficiency quantum advantage}.

\bibitem{Alcazar.15.01.2021}
\bibinfo{author}{Alcazar, J.}, \bibinfo{author}{Vakili, M.~G.},
  \bibinfo{author}{Kalayci, C.~B.} \& \bibinfo{author}{Perdomo-Ortiz, A.}
\newblock \bibinfo{title}{Geo: Enhancing combinatorial optimization with
  classical and quantum generative models}.

\bibitem{Mugel.2022}
\bibinfo{author}{Mugel, S.} \emph{et~al.}
\newblock \bibinfo{journal}{\bibinfo{title}{Dynamic portfolio optimization with
  real datasets using quantum processors and quantum-inspired tensor
  networks}}.
\newblock {\emph{\JournalTitle{Physical Review Research}}}
  \textbf{\bibinfo{volume}{4}}, \bibinfo{pages}{77},
  \doiprefix\doiurl{10.1103/PhysRevResearch.4.013006} (\bibinfo{year}{2022}).

\bibitem{Cavinato.2021}
\bibinfo{author}{Cavinato, S.}, \bibinfo{author}{Felser, T.},
  \bibinfo{author}{Fusella, M.}, \bibinfo{author}{Paiusco, M.} \&
  \bibinfo{author}{Montangero, S.}
\newblock \bibinfo{journal}{\bibinfo{title}{Optimizing radiotherapy plans for
  cancer treatment with tensor networks}}.
\newblock {\emph{\JournalTitle{Physics in medicine and biology}}}
  \textbf{\bibinfo{volume}{66}}, \doiprefix\doiurl{10.1088/1361-6560/ac01f2}
  (\bibinfo{year}{2021}).

\bibitem{Sierra.1998}
\bibinfo{author}{Sierra, G.} \& \bibinfo{author}{Martin-Delgado, M.~A.}
\newblock \bibinfo{title}{The density matrix renormalization group, quantum
  groups and conformal field theory},
  \doiprefix\doiurl{10.48550/ARXIV.COND-MAT/9811170}.

\bibitem{Uvarov.2020}
\bibinfo{author}{Uvarov, A.}, \bibinfo{author}{Kardashin, A.} \&
  \bibinfo{author}{Biamonte, J.}
\newblock \bibinfo{journal}{\bibinfo{title}{Machine learning phase transitions
  with a quantum processor}}.
\newblock {\emph{\JournalTitle{Physical Review A}}}
  \textbf{\bibinfo{volume}{102}},
  \doiprefix\doiurl{10.1103/PhysRevA.102.012415} (\bibinfo{year}{2020}).

\bibitem{Lubasch.2020}
\bibinfo{author}{Lubasch, M.}, \bibinfo{author}{Joo, J.},
  \bibinfo{author}{Moinier, P.}, \bibinfo{author}{Kiffner, M.} \&
  \bibinfo{author}{Jaksch, D.}
\newblock \bibinfo{journal}{\bibinfo{title}{Variational quantum algorithms for
  nonlinear problems}}.
\newblock {\emph{\JournalTitle{Physical Review A}}}
  \textbf{\bibinfo{volume}{101}}, \bibinfo{pages}{451},
  \doiprefix\doiurl{10.1103/PhysRevA.101.010301} (\bibinfo{year}{2020}).

\bibitem{Lazzarin.2022}
\bibinfo{author}{Lazzarin, M.}, \bibinfo{author}{Galli, D.~E.} \&
  \bibinfo{author}{Prati, E.}
\newblock \bibinfo{journal}{\bibinfo{title}{Multi-class quantum classifiers
  with tensor network circuits for quantum phase recognition}}.
\newblock {\emph{\JournalTitle{Physics Letters A}}}
  \textbf{\bibinfo{volume}{434}}, \bibinfo{pages}{128056},
  \doiprefix\doiurl{10.1016/j.physleta.2022.128056} (\bibinfo{year}{2022}).

\bibitem{Murg.2015}
\bibinfo{author}{Murg, V.}, \bibinfo{author}{Verstraete, F.},
  \bibinfo{author}{Schneider, R.}, \bibinfo{author}{Nagy, P.~R.} \&
  \bibinfo{author}{Legeza, {\"O}.}
\newblock \bibinfo{journal}{\bibinfo{title}{Tree tensor network state with
  variable tensor order: An efficient multireference method for strongly
  correlated systems}}.
\newblock {\emph{\JournalTitle{Journal of chemical theory and computation}}}
  \textbf{\bibinfo{volume}{11}}, \bibinfo{pages}{1027--1036},
  \doiprefix\doiurl{10.1021/ct501187j} (\bibinfo{year}{2015}).

\bibitem{Vidal.}
\bibinfo{author}{Vidal, G.}
\newblock \bibinfo{title}{Entanglement renormalization}.

\bibitem{Araz.21.02.2022}
\bibinfo{author}{Araz, J.~Y.} \& \bibinfo{author}{Spannowsky, M.}
\newblock \bibinfo{title}{Classical versus quantum: comparing tensor
  network-based quantum circuits on lhc data}.

\bibitem{Tagliacozzo.2009}
\bibinfo{author}{Tagliacozzo, L.}, \bibinfo{author}{Evenbly, G.} \&
  \bibinfo{author}{Vidal, G.}
\newblock \bibinfo{journal}{\bibinfo{title}{Simulation of two-dimensional
  quantum systems using a tree tensor network that exploits the entropic area
  law}}.
\newblock {\emph{\JournalTitle{Physical Review B}}}
  \textbf{\bibinfo{volume}{80}}, \doiprefix\doiurl{10.1103/PhysRevB.80.235127}
  (\bibinfo{year}{2009}).

\bibitem{Cincio.2008}
\bibinfo{author}{Cincio, L.}, \bibinfo{author}{Dziarmaga, J.} \&
  \bibinfo{author}{Rams, M.~M.}
\newblock \bibinfo{journal}{\bibinfo{title}{Multiscale entanglement
  renormalization ansatz in two dimensions: quantum ising model}}.
\newblock {\emph{\JournalTitle{Physical review letters}}}
  \textbf{\bibinfo{volume}{100}}, \bibinfo{pages}{240603},
  \doiprefix\doiurl{10.1103/PhysRevLett.100.240603} (\bibinfo{year}{2008}).

\bibitem{Wolf.2008}
\bibinfo{author}{Wolf, M.~M.}, \bibinfo{author}{Verstraete, F.},
  \bibinfo{author}{Hastings, M.~B.} \& \bibinfo{author}{Cirac, J.~I.}
\newblock \bibinfo{journal}{\bibinfo{title}{Area laws in quantum systems:
  mutual information and correlations}}.
\newblock {\emph{\JournalTitle{Physical review letters}}}
  \textbf{\bibinfo{volume}{100}}, \bibinfo{pages}{070502},
  \doiprefix\doiurl{10.1103/PhysRevLett.100.070502} (\bibinfo{year}{2008}).

\bibitem{Vidal.2004}
\bibinfo{author}{Vidal, G.}
\newblock \bibinfo{journal}{\bibinfo{title}{Efficient simulation of
  one-dimensional quantum many-body systems}}.
\newblock {\emph{\JournalTitle{Physical review letters}}}
  \textbf{\bibinfo{volume}{93}}, \bibinfo{pages}{040502},
  \doiprefix\doiurl{10.1103/PhysRevLett.93.040502} (\bibinfo{year}{2004}).

\bibitem{Daley.2004}
\bibinfo{author}{Daley, A.~J.}, \bibinfo{author}{Kollath, C.},
  \bibinfo{author}{Schollw{\"o}ck, U.} \& \bibinfo{author}{Vidal, G.}
\newblock \bibinfo{journal}{\bibinfo{title}{Time-dependent density-matrix
  renormalization-group using adaptive effective hilbert spaces}}.
\newblock {\emph{\JournalTitle{Journal of Statistical Mechanics: Theory and
  Experiment}}} \textbf{\bibinfo{volume}{2004}}, \bibinfo{pages}{P04005},
  \doiprefix\doiurl{10.1088/1742-5468/2004/04/P04005} (\bibinfo{year}{2004}).

\bibitem{Barratt.2022}
\bibinfo{author}{Barratt, F.}, \bibinfo{author}{Dborin, J.} \&
  \bibinfo{author}{Wright, L.}
\newblock \bibinfo{journal}{\bibinfo{title}{Improvements to gradient descent
  methods for quantum tensor network machine learning}}.
\newblock {\emph{\JournalTitle{Second Workshop on Quantum Tensor Networks in
  Machine Learning}}}  (\bibinfo{year}{2022}).

\bibitem{Wall.2021}
\bibinfo{author}{Wall, M.~L.} \& \bibinfo{author}{D'Aguanno, G.}
\newblock \bibinfo{journal}{\bibinfo{title}{Tree tensor network classifiers for
  machine learning: from quantum-inspired to quantum-assisted}}.
\newblock {\emph{\JournalTitle{Physical Review A}}}
  \textbf{\bibinfo{volume}{104}}, \bibinfo{pages}{1498},
  \doiprefix\doiurl{10.1103/PhysRevA.104.042408} (\bibinfo{year}{2021}).

\bibitem{Spall.1992}
\bibinfo{author}{Spall, J.~C.}
\newblock \bibinfo{journal}{\bibinfo{title}{Multivariate stochastic
  approximation using a simultaneous perturbation gradient approximation}}.
\newblock {\emph{\JournalTitle{IEEE Transactions on Automatic Control}}}
  \textbf{\bibinfo{volume}{37}}, \bibinfo{pages}{332--341},
  \doiprefix\doiurl{10.1109/9.119632} (\bibinfo{year}{1992}).

\bibitem{Grant.2018}
\bibinfo{author}{Grant, E.} \emph{et~al.}
\newblock \bibinfo{journal}{\bibinfo{title}{Hierarchical quantum classifiers}}.
\newblock {\emph{\JournalTitle{npj Quantum Information}}}
  \textbf{\bibinfo{volume}{4}}, \doiprefix\doiurl{10.1038/s41534-018-0116-9}
  (\bibinfo{year}{2018}).

\bibitem{McClean.2018}
\bibinfo{author}{McClean, J.~R.}, \bibinfo{author}{Boixo, S.},
  \bibinfo{author}{Smelyanskiy, V.~N.}, \bibinfo{author}{Babbush, R.} \&
  \bibinfo{author}{Neven, H.}
\newblock \bibinfo{journal}{\bibinfo{title}{Barren plateaus in quantum neural
  network training landscapes}}.
\newblock {\emph{\JournalTitle{Nature communications}}}
  \textbf{\bibinfo{volume}{9}}, \bibinfo{pages}{4812},
  \doiprefix\doiurl{10.1038/s41467-018-07090-4} (\bibinfo{year}{2018}).

\bibitem{Hochreiter.1991}
\bibinfo{author}{Hochreiter, J.}
\newblock \emph{\bibinfo{title}{Untersuchungen zu dynamischen neuronalen
  Netzen}} (\bibinfo{publisher}{thesis}, \bibinfo{address}{M{\"u}nchen},
  \bibinfo{year}{1991}).

\bibitem{Swingle.2012}
\bibinfo{author}{Swingle, B.}
\newblock \bibinfo{journal}{\bibinfo{title}{Entanglement renormalization and
  holography}}.
\newblock {\emph{\JournalTitle{Physical Review D}}}
  \textbf{\bibinfo{volume}{86}}, \bibinfo{pages}{231},
  \doiprefix\doiurl{10.1103/PhysRevD.86.065007} (\bibinfo{year}{2012}).

\bibitem{Rohwedder.2013}
\bibinfo{author}{Rohwedder, T.} \& \bibinfo{author}{Uschmajew, A.}
\newblock \bibinfo{journal}{\bibinfo{title}{On local convergence of alternating
  schemes for optimization of convex problems in the tensor train format}}.
\newblock {\emph{\JournalTitle{SIAM Journal on Numerical Analysis}}}
  \textbf{\bibinfo{volume}{51}}, \bibinfo{pages}{1134--1162},
  \doiprefix\doiurl{10.1137/110857520} (\bibinfo{year}{2013}).

\bibitem{Novikov.12.05.2016}
\bibinfo{author}{Novikov, A.}, \bibinfo{author}{Trofimov, M.} \&
  \bibinfo{author}{Oseledets, I.}
\newblock \bibinfo{title}{Exponential machines}.

\bibitem{Luchnikov.2021}
\bibinfo{author}{Luchnikov, I.~A.}, \bibinfo{author}{Krechetov, M.~E.} \&
  \bibinfo{author}{Filippov, S.~N.}
\newblock \bibinfo{journal}{\bibinfo{title}{Riemannian geometry and automatic
  differentiation for optimization problems of quantum physics and quantum
  technologies}}.
\newblock {\emph{\JournalTitle{New Journal of Physics}}}
  \textbf{\bibinfo{volume}{23}}, \bibinfo{pages}{073006},
  \doiprefix\doiurl{10.1088/1367-2630/ac0b02} (\bibinfo{year}{2021}).

\bibitem{Hauru.2021}
\bibinfo{author}{Hauru, M.}, \bibinfo{author}{{van Damme}, M.} \&
  \bibinfo{author}{Haegeman, J.}
\newblock \bibinfo{journal}{\bibinfo{title}{Riemannian optimization of
  isometric tensor networks}}.
\newblock {\emph{\JournalTitle{SciPost Physics}}}
  \textbf{\bibinfo{volume}{10}},
  \doiprefix\doiurl{10.21468/SciPostPhys.10.2.040} (\bibinfo{year}{2021}).

\bibitem{EdwinStoudenmire.}
\bibinfo{author}{{Edwin Stoudenmire}} \& \bibinfo{author}{{David J. Schwab}}.
\newblock \bibinfo{journal}{\bibinfo{title}{Supervised learning with tensor
  networks}}.
\newblock {\emph{\JournalTitle{30th Conference on Neural Information Processing
  Systems}}}  (\bibinfo{year}{2016}).

\bibitem{Wall.2021b}
\bibinfo{author}{Wall, M.~L.}, \bibinfo{author}{Abernathy, M.~R.} \&
  \bibinfo{author}{Quiroz, G.}
\newblock \bibinfo{journal}{\bibinfo{title}{Generative machine learning with
  tensor networks: benchmarks on near-term quantum computers}}.
\newblock {\emph{\JournalTitle{Physical Review Research}}}
  \textbf{\bibinfo{volume}{3}},
  \doiprefix\doiurl{10.1103/PhysRevResearch.3.023010} (\bibinfo{year}{2021}).

\bibitem{Glasser.15.06.2018}
\bibinfo{author}{Glasser, I.}, \bibinfo{author}{Pancotti, N.} \&
  \bibinfo{author}{Cirac, J.~I.}
\newblock \bibinfo{title}{From probabilistic graphical models to generalized
  tensor networks for supervised learning}.

\bibitem{Reyes.22.01.2020}
\bibinfo{author}{Reyes, J.} \& \bibinfo{author}{Stoudenmire, M.}
\newblock \bibinfo{title}{A multi-scale tensor network architecture for
  classification and regression}.

\bibitem{Stoudenmire.2018}
\bibinfo{author}{Stoudenmire, E.~M.}
\newblock \bibinfo{journal}{\bibinfo{title}{Learning relevant features of data
  with multi-scale tensor networks}}.
\newblock {\emph{\JournalTitle{Quantum Science and Technology}}}
  \textbf{\bibinfo{volume}{3}}, \bibinfo{pages}{034003},
  \doiprefix\doiurl{10.1088/2058-9565/aaba1a} (\bibinfo{year}{2018}).

\bibitem{Chen.2018}
\bibinfo{author}{Chen, J.}, \bibinfo{author}{Cheng, S.}, \bibinfo{author}{Xie,
  H.}, \bibinfo{author}{Wang, L.} \& \bibinfo{author}{Xiang, T.}
\newblock \bibinfo{journal}{\bibinfo{title}{Equivalence of restricted boltzmann
  machines and tensor network states}}.
\newblock {\emph{\JournalTitle{Physical Review B}}}
  \textbf{\bibinfo{volume}{97}}, \doiprefix\doiurl{10.1103/PhysRevB.97.085104}
  (\bibinfo{year}{2018}).

\bibitem{Kong.}
\bibinfo{author}{Kong, F.}, \bibinfo{author}{Liu, X.-y.} \&
  \bibinfo{author}{Henao, R.}
\newblock \bibinfo{journal}{\bibinfo{title}{Quantum tensor network in machine
  learning: An application to tiny object classification}}.
\newblock {\emph{\JournalTitle{34th Conference on Neural Information}}}
  (\bibinfo{year}{2020}).

\bibitem{Novikov.2015}
\bibinfo{author}{Novikov, A.}, \bibinfo{author}{Podoprikhin, D.},
  \bibinfo{author}{Osokin, A.} \& \bibinfo{author}{Vetrov, D.~P.}
\newblock \bibinfo{title}{Tensorizing neural networks}.
\newblock In \bibinfo{editor}{{C. Cortes}}, \bibinfo{editor}{{N. Lawrence}},
  \bibinfo{editor}{{D. Lee}}, \bibinfo{editor}{{M. Sugiyama}} \&
  \bibinfo{editor}{{R. Garnett}} (eds.) \emph{\bibinfo{booktitle}{Advances in
  Neural Information Processing Systems}}, vol.~\bibinfo{volume}{28}
  (\bibinfo{publisher}{{Curran Associates, Inc}}, \bibinfo{year}{2015}).

\bibitem{Glasser.2019}
\bibinfo{author}{Glasser, I.}, \bibinfo{author}{Sweke, R.},
  \bibinfo{author}{Pancotti, N.}, \bibinfo{author}{Eisert, J.} \&
  \bibinfo{author}{Cirac, J.~I.}
\newblock \bibinfo{journal}{\bibinfo{title}{Expressive power of tensor-network
  factorizations for probabilistic modeling, with applications from hidden
  markov models to quantum machine learning}}.
\newblock {\emph{\JournalTitle{Advances in Neural Information Processing
  Systems 32}}}  (\bibinfo{year}{2019}).

\bibitem{Li.2021}
\bibinfo{author}{Li, S.}, \bibinfo{author}{Pan, F.}, \bibinfo{author}{Zhou, P.}
  \& \bibinfo{author}{Zhang, P.}
\newblock \bibinfo{journal}{\bibinfo{title}{Boltzmann machines as
  two-dimensional tensor networks}}.
\newblock {\emph{\JournalTitle{Physical Review B}}}
  \textbf{\bibinfo{volume}{104}}, \bibinfo{pages}{21},
  \doiprefix\doiurl{10.1103/PhysRevB.104.075154} (\bibinfo{year}{2021}).

\bibitem{Collura.2021}
\bibinfo{author}{Collura, M.}, \bibinfo{author}{Dell'Anna, L.},
  \bibinfo{author}{Felser, T.} \& \bibinfo{author}{Montangero, S.}
\newblock \bibinfo{journal}{\bibinfo{title}{On the descriptive power of
  neural-networks as constrained tensor networks with exponentially large bond
  dimension}}.
\newblock {\emph{\JournalTitle{SciPost Physics Core}}}
  \textbf{\bibinfo{volume}{4}},
  \doiprefix\doiurl{10.21468/SciPostPhysCore.4.1.001} (\bibinfo{year}{2021}).

\bibitem{Wu.24.06.2022}
\bibinfo{author}{Wu, D.}, \bibinfo{author}{Rossi, R.},
  \bibinfo{author}{Vicentini, F.} \& \bibinfo{author}{Carleo, G.}
\newblock \bibinfo{title}{From tensor network quantum states to tensorial
  recurrent neural networks}.

\bibitem{Levine.2019}
\bibinfo{author}{Levine, Y.}, \bibinfo{author}{Sharir, O.},
  \bibinfo{author}{Cohen, N.} \& \bibinfo{author}{Shashua, A.}
\newblock \bibinfo{journal}{\bibinfo{title}{Quantum entanglement in deep
  learning architectures}}.
\newblock {\emph{\JournalTitle{Physical Review Letters}}}
  \textbf{\bibinfo{volume}{122}}, \bibinfo{pages}{401},
  \doiprefix\doiurl{10.1103/PhysRevLett.122.065301} (\bibinfo{year}{2019}).

\bibitem{Araz.2021}
\bibinfo{author}{Araz, J.~Y.} \& \bibinfo{author}{Spannowsky, M.}
\newblock \bibinfo{journal}{\bibinfo{title}{Quantum-inspired event
  reconstruction with tensor networks: Matrix product states}}.
\newblock {\emph{\JournalTitle{Journal of High Energy Physics}}}
  \textbf{\bibinfo{volume}{2021}}, \doiprefix\doiurl{10.1007/JHEP08(2021)112}
  (\bibinfo{year}{2021}).

\bibitem{Felser.2021}
\bibinfo{author}{Felser, T.} \emph{et~al.}
\newblock \bibinfo{journal}{\bibinfo{title}{Quantum-inspired machine learning
  on high-energy physics data}}.
\newblock {\emph{\JournalTitle{npj Quantum Information}}}
  \textbf{\bibinfo{volume}{7}}, \doiprefix\doiurl{10.1038/s41534-021-00443-w}
  (\bibinfo{year}{2021}).

\bibitem{Selvan.13.11.2020}
\bibinfo{author}{Selvan, R.}, \bibinfo{author}{{\O}rting, S.} \&
  \bibinfo{author}{Dam, E.~B.}
\newblock \bibinfo{title}{Multi-layered tensor networks for image
  classification}.

\bibitem{Liu.2021}
\bibinfo{author}{Liu, Y.} \emph{et~al.}
\newblock \bibinfo{journal}{\bibinfo{title}{Entanglement-based feature
  extraction by tensor network machine learning}}.
\newblock {\emph{\JournalTitle{Frontiers in Applied Mathematics and
  Statistics}}} \textbf{\bibinfo{volume}{7}},
  \doiprefix\doiurl{10.3389/fams.2021.716044} (\bibinfo{year}{2021}).

\bibitem{Wang.03.06.2020}
\bibinfo{author}{Wang, J.}, \bibinfo{author}{Roberts, C.},
  \bibinfo{author}{Vidal, G.} \& \bibinfo{author}{Leichenauer, S.}
\newblock \bibinfo{title}{Anomaly detection with tensor networks}.

\bibitem{Sun.2020}
\bibinfo{author}{Sun, Z.-Z.}, \bibinfo{author}{Peng, C.}, \bibinfo{author}{Liu,
  D.}, \bibinfo{author}{Ran, S.-J.} \& \bibinfo{author}{Su, G.}
\newblock \bibinfo{journal}{\bibinfo{title}{Generative tensor network
  classification model for supervised machine learning}}.
\newblock {\emph{\JournalTitle{Physical Review B}}}
  \textbf{\bibinfo{volume}{101}},
  \doiprefix\doiurl{10.1103/PhysRevB.101.075135} (\bibinfo{year}{2020}).

\bibitem{Han.2018}
\bibinfo{author}{Han, Z.-Y.}, \bibinfo{author}{Wang, J.}, \bibinfo{author}{Fan,
  H.}, \bibinfo{author}{Wang, L.} \& \bibinfo{author}{Zhang, P.}
\newblock \bibinfo{journal}{\bibinfo{title}{Unsupervised generative modeling
  using matrix product states}}.
\newblock {\emph{\JournalTitle{Physical Review X}}}
  \textbf{\bibinfo{volume}{8}}, \doiprefix\doiurl{10.1103/PhysRevX.8.031012}
  (\bibinfo{year}{2018}).

\bibitem{Cheng.2019}
\bibinfo{author}{Cheng, S.}, \bibinfo{author}{Wang, L.},
  \bibinfo{author}{Xiang, T.} \& \bibinfo{author}{Zhang, P.}
\newblock \bibinfo{journal}{\bibinfo{title}{Tree tensor networks for generative
  modeling}}.
\newblock {\emph{\JournalTitle{Physical Review B}}}
  \textbf{\bibinfo{volume}{99}}, \doiprefix\doiurl{10.1103/PhysRevB.99.155131}
  (\bibinfo{year}{2019}).

\bibitem{Bai.2022}
\bibinfo{author}{Bai, S.-C.}, \bibinfo{author}{Tang, Y.-C.} \&
  \bibinfo{author}{Ran, S.-J.}
\newblock \bibinfo{journal}{\bibinfo{title}{Unsupervised recognition of
  informative features via tensor network machine learning and quantum
  entanglement variations}}.
\newblock {\emph{\JournalTitle{Chinese Physics Letters}}}
  \textbf{\bibinfo{volume}{39}}, \bibinfo{pages}{100701},
  \doiprefix\doiurl{10.1088/0256-307X/39/10/100701} (\bibinfo{year}{2022}).

\bibitem{Liu.2019}
\bibinfo{author}{Liu, D.} \emph{et~al.}
\newblock \bibinfo{journal}{\bibinfo{title}{Machine learning by unitary tensor
  network of hierarchical tree structure}}.
\newblock {\emph{\JournalTitle{New Journal of Physics}}}
  \textbf{\bibinfo{volume}{21}}, \bibinfo{pages}{073059},
  \doiprefix\doiurl{10.1088/1367-2630/ab31ef} (\bibinfo{year}{2019}).

\bibitem{Wall.2022}
\bibinfo{author}{Wall, M.~L.}, \bibinfo{author}{Titum, P.},
  \bibinfo{author}{Quiroz, G.}, \bibinfo{author}{Foss-Feig, M.} \&
  \bibinfo{author}{Hazzard, K. R.~A.}
\newblock \bibinfo{journal}{\bibinfo{title}{A tensor network discriminator
  architecture for classification of quantum data on quantum computers}}.
\newblock {\emph{\JournalTitle{Physical Review A}}}
  \textbf{\bibinfo{volume}{105}}, \bibinfo{pages}{520},
  \doiprefix\doiurl{10.1103/PhysRevA.105.062439} (\bibinfo{year}{2022}).

\bibitem{Dborin.2022}
\bibinfo{author}{Dborin, J.}, \bibinfo{author}{Barratt, F.},
  \bibinfo{author}{Wimalaweera, V.}, \bibinfo{author}{Wright, L.} \&
  \bibinfo{author}{Green, A.~G.}
\newblock \bibinfo{journal}{\bibinfo{title}{Matrix product state pre-training
  for quantum machine learning}}.
\newblock {\emph{\JournalTitle{Quantum Science and Technology}}}
  \textbf{\bibinfo{volume}{7}}, \bibinfo{pages}{035014},
  \doiprefix\doiurl{10.1088/2058-9565/ac7073} (\bibinfo{year}{2022}).

\bibitem{Huggins.2019}
\bibinfo{author}{Huggins, W.}, \bibinfo{author}{Patil, P.},
  \bibinfo{author}{Mitchell, B.}, \bibinfo{author}{Whaley, K.~B.} \&
  \bibinfo{author}{Stoudenmire, E.~M.}
\newblock \bibinfo{journal}{\bibinfo{title}{Towards quantum machine learning
  with tensor networks}}.
\newblock {\emph{\JournalTitle{Quantum Science and Technology}}}
  \textbf{\bibinfo{volume}{4}}, \bibinfo{pages}{024001},
  \doiprefix\doiurl{10.1088/2058-9565/aaea94} (\bibinfo{year}{2019}).

\bibitem{Ran.2020}
\bibinfo{author}{Ran, S.-J.}
\newblock \bibinfo{journal}{\bibinfo{title}{Encoding of matrix product states
  into quantum circuits of one- and two-qubit gates}}.
\newblock {\emph{\JournalTitle{Physical Review A}}}
  \textbf{\bibinfo{volume}{101}}, \bibinfo{pages}{401},
  \doiprefix\doiurl{10.1103/PhysRevA.101.032310} (\bibinfo{year}{2020}).

\bibitem{Lin.2021}
\bibinfo{author}{Lin, S.-H.}, \bibinfo{author}{Dilip, R.},
  \bibinfo{author}{Green, A.~G.}, \bibinfo{author}{Smith, A.} \&
  \bibinfo{author}{Pollmann, F.}
\newblock \bibinfo{journal}{\bibinfo{title}{Real- and imaginary-time evolution
  with compressed quantum circuits}}.
\newblock {\emph{\JournalTitle{PRX Quantum}}} \textbf{\bibinfo{volume}{2}},
  \doiprefix\doiurl{10.1103/PRXQuantum.2.010342} (\bibinfo{year}{2021}).

\bibitem{Schon.2005}
\bibinfo{author}{Sch{\"o}n, C.}, \bibinfo{author}{Solano, E.},
  \bibinfo{author}{Verstraete, F.}, \bibinfo{author}{Cirac, J.~I.} \&
  \bibinfo{author}{Wolf, M.~M.}
\newblock \bibinfo{journal}{\bibinfo{title}{Sequential generation of entangled
  multiqubit states}}.
\newblock {\emph{\JournalTitle{Physical review letters}}}
  \textbf{\bibinfo{volume}{95}}, \bibinfo{pages}{110503},
  \doiprefix\doiurl{10.1103/PhysRevLett.95.110503} (\bibinfo{year}{2005}).

\bibitem{Dilip.24.04.2022}
\bibinfo{author}{Dilip, R.}, \bibinfo{author}{Liu, Y.-J.},
  \bibinfo{author}{Smith, A.} \& \bibinfo{author}{Pollmann, F.}
\newblock \bibinfo{title}{Data compression for quantum machine learning}.

\bibitem{Guala.2022}
\bibinfo{author}{Guala, D.}, \bibinfo{author}{Cruz-Rico, E.},
  \bibinfo{author}{Zhang, S.} \& \bibinfo{author}{Arrazola, J.~M.}
\newblock \bibinfo{title}{Tensor-network quantum circuits}
  (\bibinfo{year}{2022}).

\bibitem{Fastovets.01.10.201805.10.2018}
\bibinfo{author}{Fastovets, D.~V.}, \bibinfo{author}{Bogdanov, Y.},
  \bibinfo{author}{Bantysh, B.~I.} \& \bibinfo{author}{Lukichev, V.~F.}
\newblock \bibinfo{title}{Machine learning methods in quantum computing
  theory}.
\newblock In \bibinfo{editor}{Lukichev, V.~F.} \& \bibinfo{editor}{Rudenko,
  K.~V.} (eds.) \emph{\bibinfo{booktitle}{International Conference on Micro-
  and Nano-Electronics 2018}}, \bibinfo{pages}{85},
  \doiprefix\doiurl{10.1117/12.2522427} (\bibinfo{publisher}{SPIE},
  \bibinfo{year}{01.10.2018 - 05.10.2018}).

\bibitem{Cong.2019}
\bibinfo{author}{Cong, I.}, \bibinfo{author}{Choi, S.} \&
  \bibinfo{author}{Lukin, M.~D.}
\newblock \bibinfo{journal}{\bibinfo{title}{Quantum convolutional neural
  networks}}.
\newblock {\emph{\JournalTitle{Nature Physics}}} \textbf{\bibinfo{volume}{15}},
  \bibinfo{pages}{1273--1278}, \doiprefix\doiurl{10.1038/s41567-019-0648-8}
  (\bibinfo{year}{2019}).

\bibitem{Pesah.2021}
\bibinfo{author}{Pesah, A.} \emph{et~al.}
\newblock \bibinfo{journal}{\bibinfo{title}{Absence of barren plateaus in
  quantum convolutional neural networks}}.
\newblock {\emph{\JournalTitle{Physical Review X}}}
  \textbf{\bibinfo{volume}{11}}, \doiprefix\doiurl{10.1103/PhysRevX.11.041011}
  (\bibinfo{year}{2021}).

\bibitem{Du.2020}
\bibinfo{author}{Du, Y.}, \bibinfo{author}{Hsieh, M.-H.}, \bibinfo{author}{Liu,
  T.} \& \bibinfo{author}{Tao, D.}
\newblock \bibinfo{journal}{\bibinfo{title}{Expressive power of parametrized
  quantum circuits}}.
\newblock {\emph{\JournalTitle{Physical Review Research}}}
  \textbf{\bibinfo{volume}{2}},
  \doiprefix\doiurl{10.1103/PhysRevResearch.2.033125} (\bibinfo{year}{2020}).

\bibitem{Liao.02.09.2022}
\bibinfo{author}{Liao, H.}, \bibinfo{author}{Convy, I.}, \bibinfo{author}{Yang,
  Z.} \& \bibinfo{author}{Whaley, K.~B.}
\newblock \bibinfo{title}{Decohering tensor network quantum machine learning
  models}.

\bibitem{Chen.2022}
\bibinfo{author}{Chen, S. Y.-C.}, \bibinfo{author}{Huang, C.-M.},
  \bibinfo{author}{Hsing, C.-W.}, \bibinfo{author}{Goan, H.-S.} \&
  \bibinfo{author}{Kao, Y.-J.}
\newblock \bibinfo{journal}{\bibinfo{title}{Variational quantum reinforcement
  learning via evolutionary optimization}}.
\newblock {\emph{\JournalTitle{Machine Learning: Science and Technology}}}
  \textbf{\bibinfo{volume}{3}}, \bibinfo{pages}{015025},
  \doiprefix\doiurl{10.1088/2632-2153/ac4559} (\bibinfo{year}{2022}).

\bibitem{Sagingalieva.10.05.2022}
\bibinfo{author}{Sagingalieva, A.} \emph{et~al.}
\newblock \bibinfo{title}{Hyperparameter optimization of hybrid quantum neural
  networks for car classification}.

\bibitem{Zhang.12.11.2020}
\bibinfo{author}{Zhang, K.}, \bibinfo{author}{Hsieh, M.-H.},
  \bibinfo{author}{Liu, L.} \& \bibinfo{author}{Tao, D.}
\newblock \bibinfo{title}{Toward trainability of quantum neural networks}.

\bibitem{Qi.08.06.2022}
\bibinfo{author}{Qi, J.}, \bibinfo{author}{Yang, C.-H.~H.},
  \bibinfo{author}{Chen, P.-Y.} \& \bibinfo{author}{Hsieh, M.-H.}
\newblock \bibinfo{title}{Theoretical error performance analysis for
  variational quantum circuit based functional regression}.

\bibitem{Liu.2022}
\bibinfo{author}{Liu, Z.}, \bibinfo{author}{Yu, L.-W.}, \bibinfo{author}{Duan,
  L.~M.} \& \bibinfo{author}{Deng, D.-L.}
\newblock \bibinfo{journal}{\bibinfo{title}{The presence and absence of barren
  plateaus in tensor-network based machine learning}}.
\newblock {\emph{\JournalTitle{Physical Review Letters}}}
  \textbf{\bibinfo{volume}{129}}, \bibinfo{pages}{177},
  \doiprefix\doiurl{10.1103/PhysRevLett.129.270501} (\bibinfo{year}{2022}).

\bibitem{Wang.2021}
\bibinfo{author}{Wang, K.}, \bibinfo{author}{Xiao, L.}, \bibinfo{author}{Yi,
  W.}, \bibinfo{author}{Ran, S.-J.} \& \bibinfo{author}{Xue, P.}
\newblock \bibinfo{journal}{\bibinfo{title}{Experimental realization of a
  quantum image classifier via tensor-network-based machine learning}}.
\newblock {\emph{\JournalTitle{Photonics Research}}}
  \textbf{\bibinfo{volume}{9}}, \bibinfo{pages}{2332},
  \doiprefix\doiurl{10.1364/PRJ.434217} (\bibinfo{year}{2021}).

\bibitem{Bhatia.04.05.2019}
\bibinfo{author}{Bhatia, A.~S.}, \bibinfo{author}{Saggi, M.~K.},
  \bibinfo{author}{Kumar, A.} \& \bibinfo{author}{Jain, S.}
\newblock \bibinfo{title}{Matrix product state based quantum classifier}.

\bibitem{Kardashin.2021}
\bibinfo{author}{Kardashin, A.}, \bibinfo{author}{Uvarov, A.} \&
  \bibinfo{author}{Biamonte, J.}
\newblock \bibinfo{journal}{\bibinfo{title}{Quantum machine learning tensor
  network states}}.
\newblock {\emph{\JournalTitle{Frontiers in Physics}}}
  \textbf{\bibinfo{volume}{8}}, \doiprefix\doiurl{10.3389/fphy.2020.586374}
  (\bibinfo{year}{2021}).

\bibitem{Qiskit}
\bibinfo{author}{tA~v, A.} \emph{et~al.}
\newblock \bibinfo{title}{Qiskit: An open-source framework for quantum
  computing}, \doiprefix\doiurl{10.5281/zenodo.2573505} (\bibinfo{year}{2021}).

\bibitem{CirqDevelopers.2022}
\bibinfo{author}{{Cirq Developers}}.
\newblock \bibinfo{title}{Cirq}, \doiprefix\doiurl{10.5281/ZENODO.7465577}
  (\bibinfo{year}{2022}).

\bibitem{Bergholm.12.11.2018}
\bibinfo{author}{Bergholm, V.} \emph{et~al.}
\newblock \bibinfo{title}{Pennylane: Automatic differentiation of hybrid
  quantum-classical computations}.

\bibitem{Schwarz.2012}
\bibinfo{author}{Schwarz, M.}, \bibinfo{author}{Temme, K.} \&
  \bibinfo{author}{Verstraete, F.}
\newblock \bibinfo{journal}{\bibinfo{title}{Preparing projected entangled pair
  states on a quantum computer}}.
\newblock {\emph{\JournalTitle{Physical review letters}}}
  \textbf{\bibinfo{volume}{108}}, \bibinfo{pages}{110502},
  \doiprefix\doiurl{10.1103/PhysRevLett.108.110502} (\bibinfo{year}{2012}).

\bibitem{Lu.11.03.2021}
\bibinfo{author}{Lu, S.}, \bibinfo{author}{Kan{\'a}sz-Nagy, M.},
  \bibinfo{author}{Kukuljan, I.} \& \bibinfo{author}{Cirac, J.~I.}
\newblock \bibinfo{title}{Tensor networks and efficient descriptions of
  classical data}.

\bibitem{McCord.04.03.2022}
\bibinfo{author}{McCord, J.~C.} \& \bibinfo{author}{Evenbly, G.}
\newblock \bibinfo{title}{Improved wavelets for image compression from unitary
  circuits}.

\bibitem{Le.2011}
\bibinfo{author}{Le, P.~Q.}, \bibinfo{author}{Dong, F.} \&
  \bibinfo{author}{Hirota, K.}
\newblock \bibinfo{journal}{\bibinfo{title}{A flexible representation of
  quantum images for polynomial preparation, image compression, and processing
  operations}}.
\newblock {\emph{\JournalTitle{Quantum Information Processing}}}
  \textbf{\bibinfo{volume}{10}}, \bibinfo{pages}{63--84},
  \doiprefix\doiurl{10.1007/s11128-010-0177-y} (\bibinfo{year}{2011}).

\bibitem{Latorre.2005}
\bibinfo{author}{Latorre, J.~I.}
\newblock \bibinfo{title}{Image compression and entanglement},
  \doiprefix\doiurl{10.48550/ARXIV.QUANT-PH/0510031}.

\bibitem{Chen.2021}
\bibinfo{author}{Chen, S. Y.-C.}, \bibinfo{author}{Huang, C.-M.},
  \bibinfo{author}{Hsing, C.-W.} \& \bibinfo{author}{Kao, Y.-J.}
\newblock \bibinfo{journal}{\bibinfo{title}{An end-to-end trainable hybrid
  classical-quantum classifier}}.
\newblock {\emph{\JournalTitle{Machine Learning: Science and Technology}}}
  \textbf{\bibinfo{volume}{2}}, \bibinfo{pages}{045021},
  \doiprefix\doiurl{10.1088/2632-2153/ac104d} (\bibinfo{year}{2021}).

\end{thebibliography}
